\documentclass[accepted]{uai2023} 
\usepackage[american]{babel}

\usepackage{natbib} 
\usepackage{amsmath}
\usepackage{amsfonts}
\usepackage{amssymb}
\usepackage{physics}
\usepackage{mathtools}
\usepackage{booktabs}
\usepackage{xfrac}
\usepackage{setspace}
\usepackage{pgf}
\usepackage[ruled]{algorithm2e}
\usepackage{xcolor}
\usepackage[labelsep=period]{caption}
\usepackage{subcaption}
\usepackage{placeins}
\usepackage{wrapfig}
\usepackage{bm}
\usepackage{soul}
\usepackage[ragged]{sidecap}
\usepackage{multirow}

\ifdefined\nohyperref\else\ifdefined\hypersetup
  \definecolor{mydarkblue}{rgb}{0,0.08,0.45}
  \hypersetup{ %
    pdftitle={},
    pdfsubject={},
    pdfkeywords={},
    pdfborder=0 0 0,
    pdfpagemode=UseNone,
    colorlinks=true,
    linkcolor=mydarkblue,
    citecolor=mydarkblue,
    filecolor=mydarkblue,
    urlcolor=mydarkblue,
    }
  \fi
\fi

\usepackage{amsthm}

\theoremstyle{definition}
\newtheorem{definition}{Definition}
\newtheorem{assumption}{Assumption}
\theoremstyle{remark}

\DeclareMathOperator{\E}{\mathbb{E}}

\DeclareMathOperator{\diff}{\mathrm{d}\!}

\title{Partial Identification of Dose Responses with Hidden Confounders}
\author[1]{\href{mailto:myrlm@isi.edu}{Myrl~G.~Marmarelis}{}}
\author[2]{Elizabeth~Haddad}
\author[3]{Andrew~Jesson}
\author[2]{\\Neda~Jahanshad}
\author[1]{Aram~Galstyan}
\author[1]{Greg~{Ver Steeg}}

\affil[1]{USC Information Sciences Institute\\
  4676 Admiralty Way\\
  Marina del Rey, CA 90292}

\affil[2]{USC Stevens Neuroimaging and Informatics Institute\\
  4676 Admiralty Way\\
  Marina del Rey, CA 90292}

\affil[3]{University of Oxford, OATML\\
  14 Parks Road\\
  Oxford, UK OX1 3AQ}

\newcommand{\notindep}{\ensuremath{ \mathbin{\not\!\perp\!\!\!\perp} }}
\newcommand{\indep}{\ensuremath{ \mathbin{\perp\!\!\!\perp} }}

\begin{document}
\maketitle

\begin{abstract}
  Inferring causal effects of continuous-valued treatments from observational data is a crucial task promising to better inform policy- and decision-makers. 
  A critical assumption needed to identify these effects is that all confounding variables---causal parents of both the treatment and the outcome---are included as covariates. 
  Unfortunately, given observational data alone, we cannot know with certainty that this criterion is satisfied. 
  Sensitivity analyses provide principled ways to give bounds on causal estimates when confounding variables are hidden.
  While much attention is focused on sensitivity analyses for discrete-valued treatments, much less is paid to continuous-valued treatments. 
  We present novel methodology to bound both average and conditional average continuous-valued treatment-effect estimates when they cannot be point identified due to hidden confounding.
  A semi-synthetic benchmark on multiple datasets shows our method giving tighter coverage of the true dose-response curve than a recently proposed continuous sensitivity model and baselines.
  Finally, we apply our method to a real-world observational case study to demonstrate the value of identifying dose-dependent causal effects.
\end{abstract}

\section{Introduction}\label{sec:intro}

Causal inference on observational studies~\citep{ref:hill, ref:athey19} attempts to predict conclusions of alternate versions of those studies, as if they were actually properly randomized experiments.
The causal aspect is unique among inference tasks in that the goal is not prediction per se, as causal inference deals with \emph{counterfactuals}, the problem of predicting unobservables: for example, what would have been a particular patient's health outcome had she taken some medication, versus not, while keeping all else equal (\emph{ceteris paribus})?
There is quite often no way to validate the results without bringing in additional domain knowledge.
A set of putative treatments $\mathcal{T}$, often binary with a treated/untreated dichotomy, induces \emph{potential outcomes} $Y_{t\in\mathcal{T}}$. These can depend on covariates $X$ as with heterogeneous treatment effects $\E[Y_1-Y_0 \mid X]$ in the binary case. Only one outcome is ever observed: that at the assigned treatment $T$. Potential biases arise from the incomplete observation. This problem is exacerbated with more than two treatment values, especially when there are infinite possibilities, like in a continuum, e.g.\ $\mathcal{T}=[0,1]$. Unfortunately, many consequential decisions in life involve this kind of treatment: What dose of drug should I take? How much of \rule{1.25em}{0.5pt} should I eat/drink? How much exercise do I really need?



In an observational study, the direct causal link between assigned treatment $T$ and observed outcome $Y$ (also denoted as $Y_T$) can be influenced by indirect links modulated by \emph{confounding} variables. For instance, wealth is often a confounder in an individual's health outcome from diet, medication, or exercise. Wealth affects access to each of these ``treatments,'' and it also affects health through numerous other paths. Including the confounders as covariates in $X$ allows estimators to condition on them and disentangle the influences~\citep{ref:yao}.

It can be challenging to collect sufficient data, in terms of quality and quantity, on confounders in order to adjust a causal estimation to them.
Case in point, noisy observations of e.g.\ lifestyle confounders lead researchers to vacillate on the health implications of coffee \citep{ref:atroszko2019}, alcohol \citep{ref:ystrom2022}, and cheese \citep{ref:godos2020}.



\begin{figure}[!htb]\centering
  \includegraphics[width=0.95\linewidth]{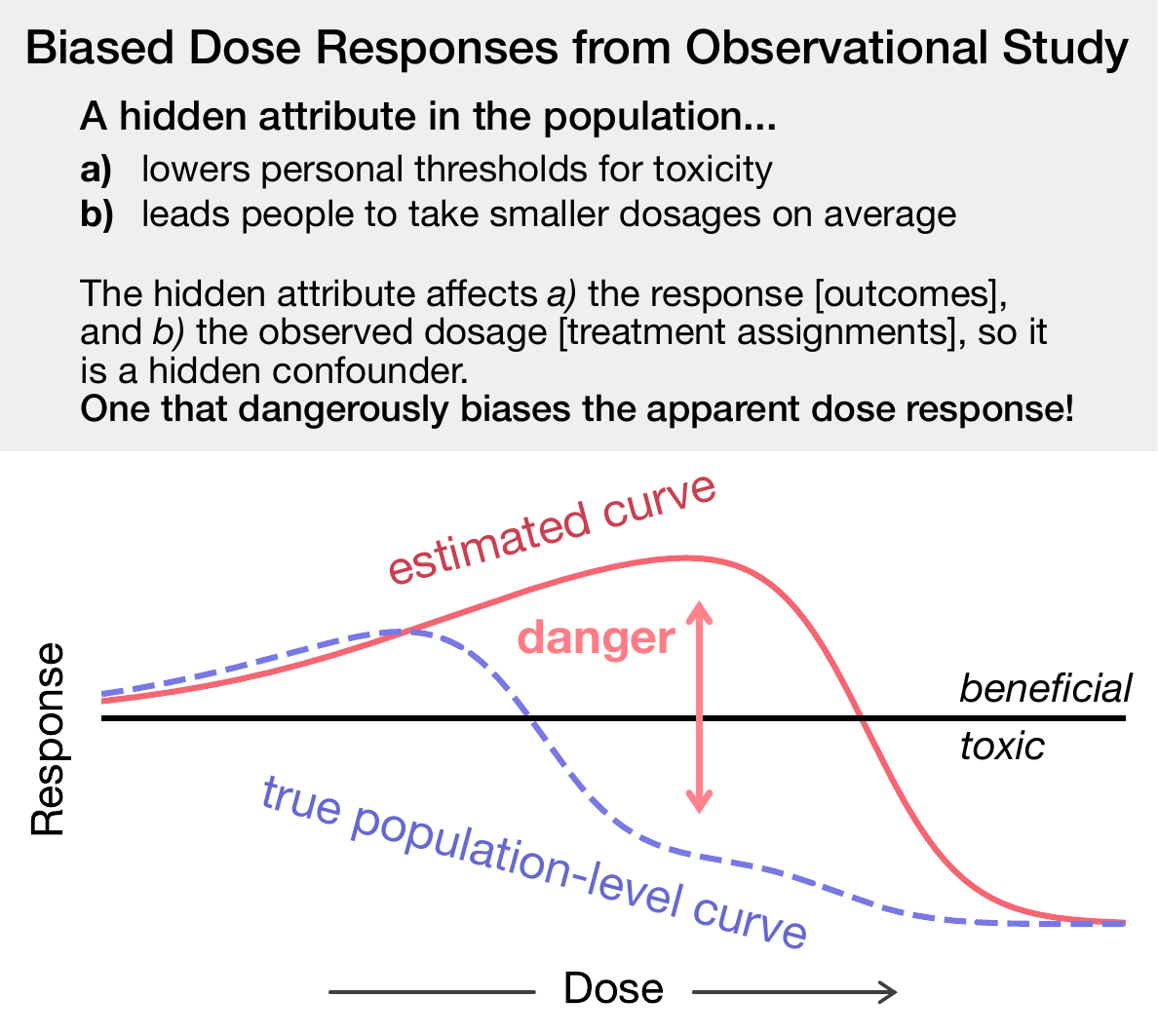}\vspace{-0.5em}
\caption{\label{fig:curve-flipping}
Dose-respone curves in medicine~\citep[e.g.][]{ref:taleb} can be viewed as expected potential outcomes from continuous treatments.
In this simulation (with details in \S D,) there is one unobserved confounder. The empirical estimate of the population-level dose responses massively overshoots the maximum effective dosage, and would suggest treatments that were actually toxic to the population. This phenomenon persists even when the vulnerable hidden subgroup occurs more often in the population. }
\end{figure}

For consequential real-world causal inference, it is only prudent to allow margins for some amount of hidden confounding. A major impediment to such analysis is that it is impossible to know how a hidden confounder would bias the causal effect. The role of any causal \emph{sensitivity model}~\citep{ref:cornfield,ref:rosenbaum83} is to make reasonable structural assumptions~\citep{ref:manski} about different levels of hidden confounding.
Most sensitivity analyses to hidden confounding require the treatment categories to be binary or at least discrete. This weakens empirical studies that are better specified by dose-response curves~\citep{ref:calabrese,ref:bonvini} from a continuous treatment variable. Estimated dose-response functions are indeed vulnerable in the presence of hidden confounders. Figure~\ref{fig:curve-flipping} highlights the danger of skewed observational studies that lead to biased estimates of personal toxic thresholds of treatment dosages. 

\subsection{Related works}\label{sec:related-works}
There is growing interest in causal methodology for continuous treatments (or exposures, interventions), especially in the fields of econometrics \citep[e.g.][]{ref:huang, ref:tubbicke}, health sciences \citep{ref:vegetabile}, and machine learning \citep{ref:chernozhukov, ref:ghassami, ref:colangelo, ref:kallus-weighting}. So far, most scrutiny on partial identification of potential outcomes has focused on the case of discrete treatments~\citep[e.g.][]{ref:rosenbaum83, ref:louizos, ref:lim}. A number of creative approaches recently made strides in the discrete setting. Most rely on a \emph{sensitivity model} for assessing the susceptibility of causal estimands to hidden-confounding bias. A sensitivity model allows hidden confounders but restricts their possible influence on the data, with an adjustable parameter that controls the overall tightness of that restriction. 

The common discrete-treatment sensitivity models are incompatible with continuous treatments, which are needed for estimating dose-response curves.
Still, some recent attempts have been made to handle hidden confounding under more general treatment domains~\citep{ref:chernozhukov}. \citet{ref:padh2022, ref:hu2021} optimize generative models to reflect bounds on the treatment effect due to ignorance, inducing an implicit sensitivity model through functional constraints. Instrumental variables are also helpful when they are available~\citep{ref:kilbertus2020}. The CMSM~\citep{ref:jesson22} was developed in parallel to this work, and now serves as a baseline.

For binary treatments, the Marginal Sensitivity Model (MSM) due to~\citet{ref:tan} has found widespread usage~\citep{ref:zhao, ref:veitch, ref:yin, ref:kallus, ref:jesson21}. Variations thereof include Rosenbaum's earlier sensitivity model~[\citeyear{ref:rosenbaum}] that enjoys ties to regression coefficients~\citep{ref:yadlowski}.
Alternatives to sensitivity models leverage generative modeling~\citep{ref:meresht} and robust optimization~\citep{ref:guo}.
Other perspectives require additional structure to the data-generating (\emph{observed outcome, treatment, covariates}) process. Proximal causal learning~\citep{ref:tchetgen, ref:mastouri} requires observation of proxy variables. \citet{ref:chen} rely on a large number of background variables to help filter out hidden confounding from apparent causal influences.

\subsection{Contributions}
We propose a novel sensitivity model for continuous treatments in \S\ref{sec:sensitivity}. Next, we derive general formulas (\S\ref{sec:deets}) and solve closed forms for three versions (\S\ref{sec:beta-weights}) of partially identified dose responses---for Beta, Gamma, and Gaussian treatment variables. We devise an efficient sampling algorithm (\S\ref{sec:optim}), and validate our results empirically using a semi-synthetic benchmark (\S\ref{sec:result-benchmark}) and realistic case study (\S\ref{sec:result-workflow}). 

\subsection{Problem Statement}
Our goal is the partial identification of causal dose responses under a bounded level of possible hidden confounding. We consider any setup that grants access to two predictors~\citep{ref:chernozhukov17} that can be learned empirically and are assumed to output correct conditional distributions. These are (1) a predictor of outcomes conditioned on covariates and the assigned treatment, and (2) a predictor of the propensity of treatment assignments, taking the form of a probability density, conditioned on the covariates. 
The latter measures (non-)uniformity in treatment assignment for different parts of the population. 
%
The observed data come from a joint distribution of outcome, continuous treatment, and covariates that include any observed confounders.

\paragraph{Potential outcomes.}
Causal inference is often cast in the nomenclature of potential outcomes, due to \citet{ref:rubin}.
Our first assumption, common to Rubin's framework, is that observation tuples of outcome, assigned treatment, and covariates, $\{(y^{(i)},t^{(i)},x^{(i)})\}_{i=1}^n,$ are \emph{i.i.d} draws from a single joint distribution. This subsumes the Stable Unit Treatment Value Assumption (SUTVA), where units/individuals cannot depend on one another, since they are \emph{i.i.d}.
The second assumption is overlap/positivity, that all treatments have a chance of assignment for every individual in the data: ${p_{T\mid X}(t\mid x)>0}$ for every ${(t,x)\in \mathcal{T}\times \mathcal{X}}$.

The third and most challenging fundamental assumption is that of \emph{ignorability}/sufficiency:
${\{(Y_t)_{t\in\mathcal{T}} \indep T\} \mid X}.$
Clearly the outcome should depend on the assigned treatment, but \emph{potential outcomes} ought not to be affected by the assignment, after blocking out paths through covariates.

Our study focuses on dealing with limited violations to ignorability.
The situation is expressed formally as 
$\{(Y_t)_{t\in\mathcal{T}} \notindep T\} \mid X$, but more specifically, we shall introduce a sensitivity model that governs the shape and extent of that violation. 

Let $p(y_t|x)$ denote the probability density function of \emph{potential} outcome $Y_t=y_t$ from a treatment $t\in \mathcal{T}$, given covariates $X=x$. This is what we seek to infer, while observing realized outcomes that allow us to learn the density $p(y_t|\,x,\,T=t)$. If the ignorability condition held, then $p(y_t|\,x,\,T=t)=p(y_t|x)$ due to the conditional independence. However, without ignorability, one has to marginalize over treatment assignment, requiring $p(y_t|\,x,\,T\neq t)$ because\vspace{-1em}
\begin{equation}\label{eq:main-integral} 
  p(y_t|x) = \int_\mathcal{T} p(y_t|\,\tau,x)p(\tau|x)\diff\tau,
\end{equation}
where $p(y_t|\tau,x)$ is the distribution of potential outcomes conditioned on actual treatment $T=\tau\in \mathcal{T}$ that may differ from the potential outcome's index $t$. The density $p(\tau|x)$ is termed the nominal propensity, defining the distribution of treatment assignments for different covariate values.


\paragraph{On notation.}
Throughout this study, $y_t$ will indicate the value of the potential outcome at treatment $t$, and to disambiguate with \emph{assigned} treatment $\tau$ will be used for events where $T=\tau$. For instance, we may care about the counterfactual of a smoker's $(\tau=1)$ health outcome had they not smoked $(y_{t=0}),$ where $T=0$ signifies no smoking and $T=1$ is ``full'' smoking. 
We will use the shorthand $p(\cdots)$ with lowercase variables whenever working with probability densities of the corresponding variables in uppercase: 
\begin{align*}
  p(y_t|\tau,x) \ &\textrm{ means } \ \frac{\partial}{\partial u}\mathbb{P}[\,Y_t\leq u\mid T=\tau,\ X=x\,]\Big\rvert_{u=y_t.}
\end{align*}


\paragraph{Quantities of interest.} We attempt to impart intuition on the conditional probability densities that may be confusing.
\begin{itemize}
  \item $p(y_t|\,x)$~~[conditional potential outcome].~~A person's outcome from a treatment, disentangled from the selection bias of treatment assignment in the population. We seek to characterize this in order to (partially) identify the Conditional Average Potential Outcome (CAPO) and the Average Potential Outcome (APO):
  \begin{equation*}
    \text{CAPO}(t, x) = \E[Y_t\mid X=x];\quad \text{APO}(t) = \E[Y_t].
  \end{equation*}
  \item $p(y_t|\,\tau,x)$~~[counterfactual].~~What is the potential outcome of a person in the population characterized by $x$ and assigned treatment $\tau$? The answer changes with $\tau$ only when $x$ is inadequate to block all backdoor paths through confounders. We can estimate this for $t=\tau$. 
  \item $p(\tau|\,y_t, x)$~~[complete propensity]~~is related to the above by Bayes' rule. We distinguish it from the nominal propensity $p(\tau|x)$ because the unobservable $y_t$ possibly confers more information about the individual, again if $x$ is inadequate. The complete propensity cannot be estimated, \emph{even for} $t=\tau$; hence, this is the target of our sensitivity model.
\end{itemize}

\begin{SCfigure}[50][ht]\centering
  \includegraphics[width=0.40\linewidth]{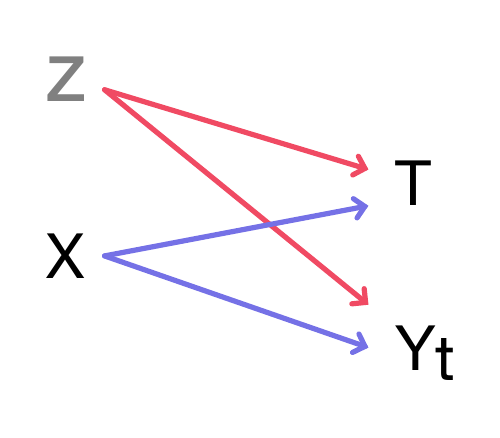}
  \caption{\label{fig:diagram}In this example, $Z$ encompasses all hidden confounders. Counterfactual $p(y_t|\,\tau,x)$ diverges from $p(y_t|\,x)$ because of the red path from $T$ to $Y_t$ through $Z$.
  }
\end{SCfigure}

A backdoor path between potential outcomes and treatment can manifest in several ways. Figure~\ref{fig:diagram} shows the barebones setting for hidden confounding to take place. Simply noisy observations of the confounders could leak a backdoor path. 
It is important to understand the ontology~\citep{ref:sarvet} of the problem in order to ascribe hidden confounding to the stochasticity inherent to a potential outcome. 

\paragraph{Sensitivity.}
Explored by \citet{ref:tan} followed by \citet{ref:kallus}, \citet{ref:jesson21}, among many others, the Marginal Sensitivity Model (MSM) serves to bound the extent of (putative) hidden confounding in the regime of binary treatments $T'\in\{0,1\}$. The MSM limits the discrepancy between the odds of treatment under the nominal propensity and the odds of treatment under the complete propensity.

\begin{definition}[The Marginal Sensitivity Model]\label{def:msm}
  For binary treatment $t'\in\{0,1\}$ and violation factor $\Gamma\geq1$, the following ratio is bounded:
  \begin{equation*}
    \Gamma^{-1}\leq\left[\frac{p(t'|x)}{1-p(t'|x)}\right]^{\,-1}\left[\frac{p(t'|\,y_{t'},x)}{1-p(t'|\,y_{t'},x)}\right]\leq\Gamma.
  \end{equation*}
\end{definition}

The confines of a binary treatment afford a number of conveniences. For instance, one probability value is sufficient to describe the whole propensity landscape on a set of conditions, $p(1-t'|\cdots)=1-p(t'|\cdots)$. As we transfer to the separate context of treatment continua, we must contend with infinite treatments and infinite potential outcomes. 

\section{Continuous Sensitivity Model}\label{sec:sensitivity}
The counterfactuals required for Equation~\ref{eq:main-integral} are almost entirely unobservable. We look to the Radon-Nikodym derivative $\omega_\delta$ of a counterfactual with respect to another~\citep{ref:tan}, quantifying their divergence between nearby treatment assignments: (assuming mutual continuity)
\begin{align*} 
  \omega_\delta(y_t|\,\tau,x) \coloneqq&\ \frac{p(y_t|\,\tau+\delta,x)}{p(y_t|\,\tau,x)}\ =\ \ \stackrel{\text{(Bayes' rule)}}{\frac{p(\tau+\delta|\,y_t, x)p(\tau|x)}{p(\tau|\,y_t, x)p(\tau+\delta|x)}}\\
  &= \left[\frac{p(\tau+\delta|x)}{p(\tau|x)}\right]^{\,-1}\left[\frac{p(\tau+\delta|\,y_t,x)}{p(\tau|\,y_t,x)}\right].
\end{align*}
As with the MSM, we encounter a ratio of odds, here contrasting $\tau$ versus $\tau+\delta$ in the assigned-treatment continuum. Assuming the densities are at least once differentiable,
\begin{equation*}
\lim_{\delta\to 0} \delta^{-1}\log\omega_\delta(y_t|\,\tau,x) = \partial_\tau [ \log p(\tau|\,y_t,x) - \log p(\tau|x)].
\end{equation*}
By constraining $\omega_\delta$ to be close to unit, via bounds above and below, we tie the logarithmic derivatives of the nominal- and complete-propensity densities.

\begin{definition}[The Infinitesimal Marginal Sensitivity Model]\label{def:lsm}
  For treatments $t\in\mathcal{T}\subseteq\mathbb{R}$, where $\mathcal{T}$ is connected, and violation-of-ignorability factor $\Gamma\geq 1$, the $\delta$MSM requires
  \begin{equation*}
    \abs{\frac{\partial}{\partial\tau} \log \frac{p(\tau|\,y_t,x)}{p(\tau|x)}} \leq \log\Gamma
  \end{equation*} 
  everywhere, for all $\tau$, $t$, and $x$ combinations. This differs from the CMSM due to \citet{ref:jesson22} that considers only $t=\tau,$ and which bounds the density ratios directly.
\end{definition}
\subsection{The Complete Framework}\label{sec:deets}

\begin{assumption}[Bounded Hidden Confounding]\label{ass:lsm}
  Invoking Definition~\ref{def:lsm}, the violation of ignorability is constrained by a $\delta$MSM with some $\Gamma\geq 1$. 
\end{assumption}
\begin{assumption}[Anchor Point]\label{ass:zero}
  A special treatment value designated as zero is not informed by potential outcomes: $p(\tau=0 \mid y_t,x)=p(\tau=0 \mid x)$ for all $x$, $t$, and $y_t$.
\end{assumption} 

At this point we state the core sensitivity assumptions. 
In addition to the $\delta$MSM, we require an anchor point at $T=0$, which may be considered a lack of treatment. Strictly, we assume that hidden confounding does not affect the propensity density precisely at the anchor point. A broader interpretation is that the strength of causal effect, hence vulnerability to hidden confounders, roughly increases with $\abs{T}$. Assumption~\ref{ass:zero} is necessary to make closed-form solutions feasible. 
We discuss ramifications and a relaxation in \S\ref{sec:beta-weights}.

\begin{figure}[hb]\centering
  \includegraphics[width=0.95\linewidth]{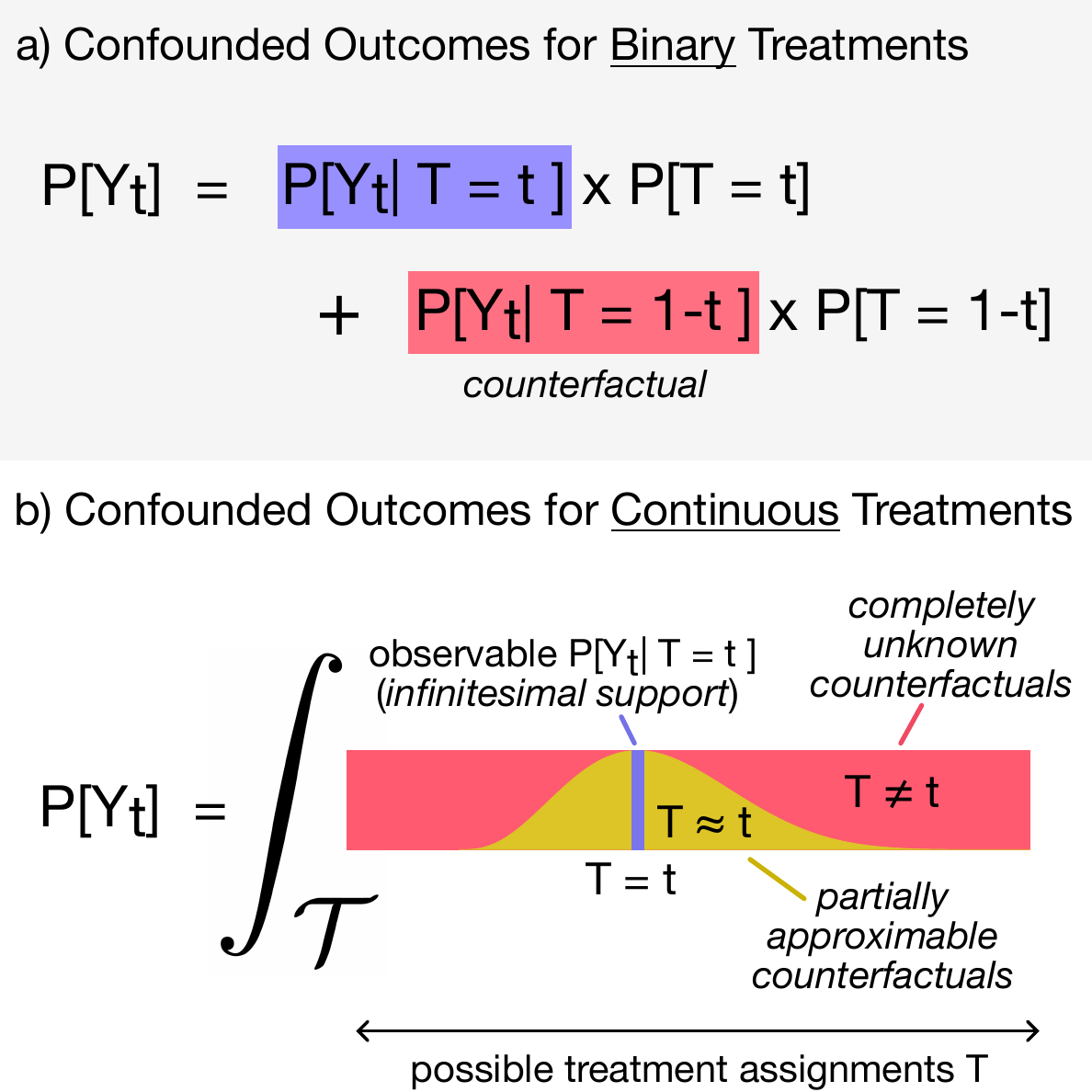}
  \caption{\label{fig:math-outline}
  In the binary case, the red part is unobservable, but the MSM condition helps to bound that quantity. In the continuous case the integrand (Equation~\ref{eq:main-integral}) is unobservable \emph{almost everywhere} in the space of assigned treatments, except for the infinitesimal point $T=t$. In order to divide the integral into two parts (observable and unobservable) like with the binary sum, we must draw an approximation where assigned treatment and potential-outcome index are close enough. We use a soft window (yellow) to mark the validity of the approximation. Our continuous version of the MSM, the $\delta$MSM, allows us to bound the red part as well as reason about the yellow part. Covariates $X$ are omitted for brevity.}
\end{figure}

The unobservability of almost all counterfactuals is unique to the case of continuous treatments, since the discrete analogy would be a discrete sum with an observable term. Figure~\ref{fig:math-outline} explains our approach to solving Equation~\ref{eq:main-integral}.


\subsection{A Partial Approximation}\label{sec:approx} 
We expand $p(y_t|\tau,x)$ around $\tau=t,$ where $p(y_t|t,x)=p(y|t,x)$ is learnable from data. Suppose that $p(y_t|\tau,x)$ is twice differentiable in $\tau$. Construct a Taylor expansion
\begin{multline}\label{eq:approx}
  p(y_t|\tau,x) = p(y_t|t,x) + (\tau-t)\partial_\tau p(y_t|\tau,x)|_{\tau=t} \\+ \frac{(\tau-t)^2}{2}\partial^2_\tau p(y_t|\tau,x)|_{\tau=t} + \mathcal{O}(\tau-t)^3.
\end{multline}
Denote with $\tilde p(y_t|\tau,x)$ an approximation of second order as laid out above. One could have stopped at lower orders but the difference in complexity is not that large. The intractable derivatives like $\partial_\tau p(y_t|\tau,x)|_{\tau=t}$ will be bounded using the $\delta$MSM machinery. Let us quantify the reliability of this approximation by a trust-weighing scheme $0\leq w_t(\tau)\leq 1,$ where typically $w_t(t)=1.$ This corresponds to the yellow part in Figure~\ref{fig:math-outline}. We argue that $w_t(\tau)$ should be narrower with lower-entropy (narrower) propensities (\S B). The possible forms of $w_t(\tau)$ are elaborated in \S\ref{sec:beta-weights}.


Splitting Equation~\ref{eq:main-integral} along the trusted regime marked by $w_t(\tau)$, and then applying the approximation of Equation~\ref{eq:approx}, 
\begin{equation}\begin{aligned}\label{eq:decompose}
  p(y_t|x) =& \int_\mathcal{T} \underbrace{w_t(\tau) p(y_t|\tau,x)p(\tau|x)\diff\tau}_\text{``observable'' (Fig.~\ref{fig:math-outline})}
  \\&+ \int_\mathcal{T} \underbrace{[1-w_t(\tau)] p(y_t|\tau,x)p(\tau|x)\diff\tau}_\text{``unobservable'' (Fig.~\ref{fig:math-outline})}\\  
  \textcolor{red!90!black}{\approx}& \int_\mathcal{T}\, \underbrace{ w_t(\tau) \textcolor{red!90!black}{\tilde p}(y_t|\tau,x)p(\tau|x)\diff\tau }_{(A)\text{ the approximated quantity}}
  \\&+ \int_\mathcal{T}\, \underbrace{ [1-w_t(\tau)] p(\tau|y_t,x)p(y_t|x)\diff\tau }_{(B)\text{ by Bayes' rule}}.
\end{aligned}\end{equation}
The intuition behind separating the integral into two parts is the following. By choosing the weights $w_t(\tau)$ so that they are close to one in the range where approximation Equation~\ref{eq:approx} is valid (yellow region in Figure~\ref{fig:math-outline}) and zero outside of this range, we can evaluate the first integral through the approximated counterfactuals. The second integral, which is effectively over the red region in Figure~\ref{fig:math-outline} and cannot be evaluated due to unobserved counterfactuals, will be bounded using the $\delta$MSM. Simplifying the second integral first,
\begin{multline*} 
  \int_\mathcal{T} [1-w_t(\tau)] p(\tau|\,y_t,x)p(y_t|x)\diff\tau \\= p(y_t|x)\left[1 - \int_\mathcal{T} w_t(\tau) p(\tau|\,y_t,x)\diff\tau \right].
\end{multline*} 
By algebraic manipulation, we witness already that $p(y_t|x)$ shall take the form of
\begin{equation}\label{eq:approx-frac}
  p(y_t|x) \approx \frac{\int_\mathcal{T} w_t(\tau) \tilde p(y_t|\,\tau,x)p(\tau|x)\diff\tau}{\int_\mathcal{T} w_t(\tau) p(\tau|\,y_t,x)\diff\tau}.
\end{equation}
Reflecting on Assumptions \ref{ass:lsm}~\&~\ref{ass:zero}, the divergence between $p(\tau|\,y_t,x)$ and $p(\tau|x)$ is bounded, allowing characterization of the denominator in terms of the learnable $p(\tau|x)$. Similarly the derivatives in Equation~\ref{eq:approx} can be bounded. These results would be sufficient to partially identify the numerator. Without loss of generality, consider the unknown quantity $\gamma$ that can be a function of $\tau$, $y_t$, and $x$, such that
\begin{multline}\label{eq:small-gamma}
  \partial_\tau \log p(\tau|y_t,x) = \partial_\tau \log p(\tau|x) + \gamma(\tau|y_t,x), \\ \text{where } \abs{\gamma(\tau|y_t,x)} \leq \log \Gamma\text{ using the $\delta$MSM.}
\end{multline}
We may attempt to integrate both sides;
\begin{align}
   \int_0^{\,s}\partial_\tau \log p(\tau|\,y_t,x)\diff\tau
     =& \int_0^{\,s}\partial_\tau \log p(\tau|x)\diff\tau \notag\\
     &\ \ + \underbrace{\int_0^{\,s}\gamma(\tau|y_t,x)\diff\tau}_{\coloneqq\lambda(s|y_t,x)}. \notag\\
  \therefore \log p(\tau=s|\,y_t,x) - \log &\, p(\tau=0|\,y_t,x) \notag\\
  = \log p(\tau=s|\,x) -& \log p(\tau=0|\,x) + \lambda(s|y_t,x), \notag\\
  \therefore \log p(\tau|\,y_t,x) = \log p(\tau|&x) + \lambda(\tau|y_t,x). \notag\\
  \textrm{(by Assumption~\ref{ass:zero})}& \notag\\
\label{eq:big-lambda} 
   \therefore \quad p(\tau|\,y_t,x) = p(\tau|x)\Lambda&(\tau|y_t,x),\ \ \Lambda \coloneqq \exp{\lambda}.
\end{align}
One finds that $\abs{\lambda(\tau|y_t,x)} \leq \abs{\tau}\log\Gamma$ because $\lambda$ integrates $\gamma$, bounded by $\pm\log\Gamma,$ over a support with length $\tau$. Subsequently, $\Lambda$ is bounded by $\Gamma^{\pm\abs{\tau}}.$ These are the requisite tools for bounding $p(y_t|x)$---or an approximation thereof, erring on ignorance via the trusted regime marked by $w_t(\tau).$ The derivation is completed in \S A by framing the unknown quantities in terms of $\gamma$ and $\Lambda$, culminating in Equation~\ref{eq:expectation}. 

\paragraph{Predicting potential outcomes.}
The recovery of a fully normalized probability density $\tilde p(y_t|x)$ via Equation~\ref{eq:approx-frac} is laid out below. It may be approximated with Monte Carlo or solved in closed form with specific formulations for the weights and propensity. Concretely, it takes on the form $\tilde p(y_t|x) = d(t|y_t,x)^{-1}p(y_t|t,x),$ where
\begin{multline}\label{eq:expectation}
  d(t|y_t,x) \coloneqq \E_\tau[\Lambda(\tau|y_t,x)]
    - [\gamma\Lambda](t|y_t,x)\,\E_\tau[\tau-t] \\
    - \frac{1}{2}[(\dot\gamma+\gamma^2)\Lambda](t|y_t,x)\,\E_\tau[(\tau-t)^2],
\end{multline}
and said expectations, $\E_\tau[\cdot],$ are with respect to the implicit distribution $q(\tau|t,x)\propto w_t(\tau)p(\tau|x).$ The notation $\dot\gamma$ denotes a derivative in the first argument of $\gamma(t|y_t,x).$

\begin{assumption}[Second-order Simplification]\label{ass:second-order}
  The quantity $\dot\gamma(\tau|y_t,x)$ cannot be characterized as-is. Granting that $\gamma^2$ dominates over the former, and consequently
  \begin{equation*}
    \abs{(\dot\gamma+\gamma^2)\Lambda} \leq \abs{\gamma^2\Lambda} + \varepsilon\quad\textrm{for small }\varepsilon \geq 0.
  \end{equation*}
\end{assumption}


\renewcommand{\arraystretch}{1.1}
\begin{table*}[bt]\centering
  \begin{tabular}{l c c c c }
    \toprule
    Parametrization & Support $(\mathcal{T})$ & Params. & Precision ($r$) & Bounds for $\E_\tau[\Lambda(\tau|y_t,x)]$ \\
    \midrule
    \textsf{Beta} & $[0,1]$ & $\alpha, \beta$ & $\alpha+\beta-2$ & ${}_1F_1(\bm\alpha+1;\ \bm\alpha+\bm\beta+2;\ \pm\log\Gamma)$ \\
    & & & & where $\bm\alpha\coloneqq \bar\alpha+\alpha-2,\ \bm\beta\coloneqq \bar\beta+\beta-2$\vspace{1em}\\
    \textsf{Balanced Beta} & $[0,1]$ & $\alpha, \beta$ & $\alpha+\beta-2$ & $t\cdot\langle\textrm{the \textsf{Beta} above}\rangle + (1-t)\cdot\langle\textrm{\textsf{Beta}, mirrored}\rangle$\vspace{1em}\\
    \textsf{Gamma} & $[0,+\infty)$ & $\alpha, \beta$ & $\alpha/\beta^2$ & $[1 - (\pm\log\Gamma)/\bm\beta]^{-\bm\alpha}$ \\
    & & & & where $\bm\alpha\coloneqq \bar\alpha+\alpha-1,\ \bm\beta\coloneqq \bar\beta+\beta$\vspace{1em}\\
    \textsf{Gaussian} & $(-\infty,+\infty)$ & $\mu, \sigma$ & $1/\sigma$ & $\exp{\bm\sigma^2(\log\Gamma)^2/2}\left(\begin{aligned}
      \Gamma^{\pm\bm\mu}[1+\erf&(\textstyle\frac{\bm\mu\pm\bm\sigma^2\log\Gamma}{\sqrt{2}\bm\sigma})] \\
      +& \\
      \Gamma^{\mp\bm\mu}[1-\erf&(\textstyle\frac{\bm\mu\mp\bm\sigma^2\log\Gamma}{\sqrt{2}\bm\sigma})]
      \end{aligned}\right)$ \vspace{0.4em}\\
      & & & & where $\bm\mu\coloneqq\frac{\mu\bar\sigma^2+\bar\mu\sigma^2}{\bar\sigma^2+\sigma^2},\ \bm\sigma^2\coloneqq\frac{\bar\sigma^2\sigma^2}{\bar\sigma^2+\sigma^2}$ \\
    \bottomrule
  \end{tabular}
  \caption{\label{tab:trusts}Candidates for propensity and trust-weighing combinations. Each row specifies the distribution---beta, beta, gamma, and Gaussian respectively---of the propensity model $p(\tau|x).$ 
  The last column lists solutions for the first term of Equation~\ref{eq:expectation}~/~\ref{eq:expectation-simplified}.
  This is a convolution of the propensity and weighing scheme, which have similar forms (see \citet{ref:bromiley} for the {\sf Gaussian} case.)
  We distinguish the replicated parameters between propensity and weight by placing a bar over the propensity parameters. So if the propensity is $x\mapsto(\bar\alpha, \bar\beta)$, then the weighing scheme has $t\mapsto(\alpha, \beta).$ The bold parameters are of the compound density, with respect to which the first and second moments are computed in Equation~\ref{eq:expectation}~/~\ref{eq:expectation-simplified}.}
\end{table*} 
\renewcommand{\arraystretch}{1}

To make use of the formula in Equation~\ref{eq:expectation}, one first obtains the set of admissible $d(t|y_t,x)\in\big[\,\underline{d}(t|y_t,x), \overline{d}(t|y_t,x)\,\big]$ that violate ignorability up to a factor $\Gamma$ according to the $\delta$MSM. With the negative side of the $\pm$ corresponding to $\underline{d}$ and the positive side to $\overline{d}$, the bounds are expressible as
\begin{multline}\label{eq:expectation-simplified}
  \Big(\underline{d},\,\overline{d}\Big)=\int_\mathcal{T} \Gamma^{\pm\abs{\tau}}q(\tau|t,x)\diff\tau \quad \text{\textcolor{red!90!black}{$\longrightarrow\E_\tau[\Lambda(\tau|y_t,x)]$}} \\
    + (\pm\log\Gamma)\Gamma^\abs{t} \,\abs{\int_\mathcal{T} (\tau-t)q(\tau|t,x)\diff\tau} \\
    + \frac{1}{2}\Big(0,\ \log^2\Gamma\Big)\Gamma^\abs{t}\,\int_\mathcal{T} (\tau-t)^2 q(\tau|t,x)\diff\tau.
\end{multline}
The $\Gamma^{\pm\abs{\tau}}$ in the first integral, as well as the alternating sign of the other two terms taken together, reveal that $\underline{d}\leq 1 \leq \overline{d}$ with equality at $\Gamma=1$. This is noteworthy because it implies that $p(y|t,x)$ is admissible for the partially identified $\tilde p(y_t|x).$ We cannot describe $p(y_t|x)$ once $\underline{d}$ crosses zero.

\paragraph{Ensembles.}
To quantify empirical uncertainties~\citep{ref:jesson20} alongside our sensitivity, the predictors could be learned as ensembles, with $\tilde p(y_t|x)$ computed as (bootstrap resampled~\citep{ref:lo}) expectations over them. 

\subsection{Propensity-Trust Combinations}\label{sec:beta-weights} 
In addition to developing the general framework above, we derive analytical forms for a myriad of paramametrizations that span the relevant supports $\mathcal{T}$ for continuous treatments: the unit interval $[0,1]$, the nonnegative reals $[0,+\infty)$, and the real number line $(-\infty,+\infty).$ For some nominal propensity distributions $p(\tau|x),$ we propose trust-weighing schemes $w_t(\tau)$ with shared form so that the expectations in Equation~\ref{eq:expectation-simplified} are solvable.

For instance, consider the parametrization $(T\mid X=x) \sim \textrm{Beta}(\alpha(x), \beta(x))$. We select a Beta-like weighing scheme, rescaled and translated, $w^\text{beta}_t(\tau) = c_t \tau^{a_t-1} (1-\tau)^{b_t-1}$. Two constraints are imposed on every $w_t(\tau)$ studied herein:
\begin{itemize}\vspace{-0.5em}
  \item \emph{(the mode)} that $w_t(\tau)$ peaks at $\tau=t$, and $w_t(t)=1$.
  \item \emph{(the precision)} that some $r>0$ defines a narrowness of the form, and can be set a priori.
\end{itemize} 

For the beta version we chose $a_t+b_t=r+2.$ These constraints imply that $a_t\coloneqq rt+1$, $b_t\coloneqq r(1-t)+1$, and $c^{-1}_t\coloneqq t^{rt}(1-t)^{r(1-t)}$.


\begin{figure}[!ht]\centering
  \scalebox{0.85}{
    \input{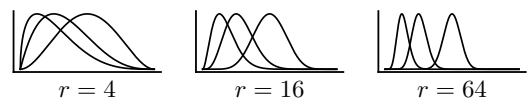}}\vspace{-1em}
  \caption{\label{fig:weights}Beta parametrizations for $w_t(\tau)$ in the unit square, plotted for $t=0.125, 0.25, 0.5$. Trust declines with $r$.}
\end{figure}

\paragraph{The choices.}
We present solutions for propensity-trust combinations in Table~\ref{tab:trusts}. {\sf Balanced Beta} stands out by not strictly obeying Assumption~\ref{ass:zero}. Rather, it adheres to a symmetrified mixture that is more versatile to realistic situations. 

\paragraph{\textsf{\textbf{Balanced Beta}}.}
  Formally, for all $t$, $y_t$, and $x$, we balance the {\sf Beta} parametrization by replacing Assumption~\ref{ass:zero} with

  $\begin{cases}
    \quad p(\,\tau=0 \mid y_t,x)=p(\,\tau=0 \mid x) &\quad \text{w.p.}\quad t,\\
    \quad p(\,\tau=1 \mid y_t,x)=p(\,\tau=1 \mid x) &\quad \text{w.p.}\quad 1-t.
  \end{cases}$

  This special parametrization deserves further justifying.
  The premise is that distant treatments are decoupled; treatment assignment $\tau$ shares less information with a distal potential outcome $y_t$ than a proximal one. 
  If that were the case, then the above linear interpolation favors the less informative anchor points for a given $t$. This is helpful because the sensitivity analysis is vulnerable to the anchor points.
  Stratifying the anchor points eventually leads to an arithmetic mixture of $d(t|y_t,x)$ in Equation~\ref{eq:expectation} with its mirrored version about $t\mapsto1-t,$ and $(\alpha,\beta)\mapsto (\beta,\alpha).$

\paragraph{Controlling trust.}
The absolute error of the approximation in Equation~\ref{eq:decompose}.A is bounded above by a form that could grow with narrower propensities (see \S B), in the Beta parametrization.
Intuitively the error also depends on the smoothness of the complete propensity (Taylor residual.) For that reason we used the heuristic of setting the trust-weighing precision $r$ to the nominal propensity precision.
\section{Estimating The Intervals}\label{sec:optim}
We seek to bound partially identified expectations with respect to the true potential-outcome densities, which are constrained according to Equation~\ref{eq:expectation}~/~\ref{eq:expectation-simplified}. The quantities of interest are the Average Potential Outcome (APO), $\E[f(Y_t)]$, and Conditional Average Potential Outcome (CAPO), $\E[f(Y_t)|X=x]$, for any task-specific $f(y)$. We use Monte Carlo over $m$ realizations $y_i$ drawn from proposal density $g(y)$, and covariates from a subsample of instances: 
\begin{multline}\label{eq:importance-sampling}
\tilde \E[f(Y_t)\mid X\in\{x^{(j)}\}_{j\in J}] = \\ \frac{\sum_{i=1}^m \sum_{j\in J} f(y_i)\, \tilde p(y_t=y_i\mid x^{(j)})/g(y_i)}{\sum_{i=1}^m \sum_{j\in J} \tilde p(y_t=y_i\mid x^{(j)})/g(y_i)},
\end{multline}
where $J\subseteq \{1\dots n\}$ indexes a subset of the finite instances. $\abs{J}=1$ recovers the formula for the CAPO, and $\abs{J}=n$ for the APO. 
The partially identified $\tilde p(y_t|x)$ really encompasses a set of probability densities that includes $p(y|t,x)$ and smooth deviations from it. Our importance sampler ensures normalization~\citep{ref:tokdar}, but is overly conservative~\citep{ref:dorn22}.
For current purposes, a greedy algorithm may be deployed to maximize (or minimize) Equation~\ref{eq:importance-sampling} by optimizing the weights $w_i$ attached to each $f(y_i)$, within the range
\begin{equation*}
  \underline{w}_i \coloneqq \frac{p(y_i|t,x)}{\overline{d}(t|y_i,x)g(y_i)},\qquad \overline{w}_i \coloneqq \frac{p(y_i|t,x)}{\underline{d}(t|y_i,x)g(y_i)}.
\end{equation*}
Our Algorithm~\ref{alg:minimax} adapts the method of \citet{ref:jesson21, ref:kallus} to heterogeneous weight bounds $[\underline{w}_i, \overline{w}_i]$ per draw $i$. View a proof of correctness in \S C.

Others have framed the APO as the averaged CAPOs, and left the min/max optimizations on the CAPO level~\citep{ref:jesson22}. We optimize the APO directly, but have not studied the impact of one choice versus the other.

\begin{algorithm}
  \SetKwInOut{Input}{Input}\SetKwInOut{Output}{Output}
  \caption{The expectation maximizer, with $\mathcal{O}(n)$ runtime if intermediate $\Delta_j$ results are memoized. \label{alg:minimax}}
  \Input{$\{(\underline{w}_i, \overline{w}_i, f_i)\}_{i=1}^n$ ordered by ascending $f_i$.}
  \Output{$\max_{w}\E[f(X)]$ estimated by importance sampling with $n$ draws.}
  \vspace{0.4em}
  Initialize $w_i \gets \overline{w}_i$ for all $i=1,2,\dots n$\;
  \For{$j=1,2,\dots n$}{
    Compute $\Delta_j\coloneqq \sum_{i=1}^n w_i(f_j-f_i)$\;
    \eIf{$\Delta_j<0$}{
      $w_j\gets \underline{w}_j$\;
    }{
      break\;
    }
  }
  Return $\sum_i w_i f_i / \sum_i w_i$
\end{algorithm}

\begin{table*}[bt]\centering
  \begin{tabular}{l| c c | c c | c c | c c | r r }
    \toprule
    Benchmarks & \multicolumn{2}{c}{\tt brain} & \multicolumn{2}{c}{\tt blood} & \multicolumn{2}{c}{\tt pbmc} & \multicolumn{2}{c|}{\tt mftc} & & \multicolumn{1}{c}{ratio} \\

    & mean & \multicolumn{1}{c}{(std.)} & mean & \multicolumn{1}{c}{(std.)} & mean & \multicolumn{1}{c}{(std.)} & mean & \multicolumn{1}{c|}{(std.)} & \% best & \multicolumn{1}{c}{to best} \\
    \midrule
    $\delta$MSM (ours) & $\bm{138}$ & $(120)$ & $\bm{141}$ & $(129)$ & $\bm{138}$ & $(121)$ & $\bm{144}$ & $(124)$ & $\bm{78.4}$ & $\bm{1.03}~(0.08)$ \\
    CMSM & $186$ & $(153)$ & $188$ & $(156)$ & $205$ & $(169)$ & $182$ & $(145)$ & $7.8$ & $1.81~(2.15)$ \\
    uniform & $158$ & $(137)$ & $162$ & $(146)$ & $157$ & $(136)$ & $167$ & $(141)$ & $4.8$ & $1.20~(0.10)$ \\
    binary MSM & $211$ & $(128)$ & $213$ & $(131)$ & $222$ & $(127)$ & $214$ & $(127)$ & $9.0$ & $2.57~(2.34)$ \\
    \bottomrule
  \end{tabular}
  \caption{\label{tab:benchmark}Semi-synthetic benchmark: divergence \underline{costs} of 90\% coverage of the Average Potential Outcome (APO), multiplied by $1000$. The four \texttt{datasets} are listed on top. We report averages over 500 trials per experiment. A paired $t$-test and sign test, roughly corresponding to the mean and median, showed significant improvement by the $\delta$MSM over the others with all $P < 10^{-5}.$
  ``\% best'' counts the proportion of trials that each method outperformed the rest, and ``ratio to best'' is the average cost ratio to the best method's in each trial---closer to one is better.}
\end{table*}

\section{A Semi-synthetic Benchmark}\label{sec:result-benchmark}
It is common practice to test causal methods, especially under novel settings, with real datasets but synthetic outcomes~\citep{ref:curth, ref:cristali}. We adopted four exceedingly diverse datasets spanning health, bioinformatics, and social-science sources. Our variable-generating process preserved the statistical idiosyncracies of each dataset. Confounders and treatment were random projections of the data, which were quantile-normalized for uniform marginals in the unit interval.
Half the confounders were observed as covariates and the other half were hidden.
The outcome was Bernoulli with random linear or quadratic forms mixing the variables before passing through a normal CDF activation function. 
Outcome and propensity models were linear and estimated by maximum likelihood.
See \S E.

\paragraph{Selecting the baselines.}
The $\delta$MSM with \textsf{Balanced Beta} was benchmarked against three relevant baselines.
\begin{itemize}
  \itemsep0.25em
  \item (CMSM)~~Use the recent model by \citet{ref:jesson22}, where $\underline{d}\coloneqq \Gamma^{-1}p(\tau|x),\ \overline{d}\coloneqq \Gamma^{+1}p(\tau|x)$.
  \item (uniform)~~Suppose $\underline{d}\coloneqq \Gamma^{-1},\ \overline{d}\coloneqq \Gamma^{+1}$, as if the propensity were uniform and constant.
  \item (binary MSM)~~Shoehorn the propensity into the classic MSM~\citep{ref:tan} by considering the treatment as binary with indicator $\mathbb{I}[T>0.5].$ 
\end{itemize} 
Note that the CMSM becomes equivalent to the ``uniform'' baseline above when CAPOs are concerned (Equation~\ref{eq:importance-sampling} with $m=1$), which are not studied in this benchmark.

\begin{figure}[!ht]\centering
  \scalebox{0.75}{
    \input{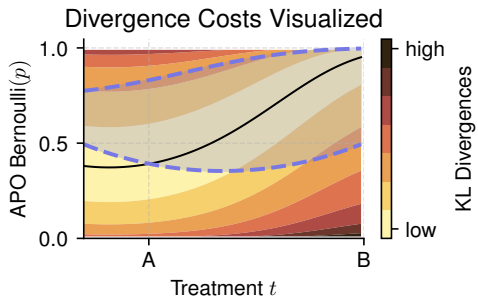}}\vspace{-1em}
  \caption{\label{fig:divergence-cost}Divergence cost measures the size of the ignorance intervals (blue), weighted by the badness of each estimate (red). The black line is the true APO. Coverage is the portion of treatments contained in the blue shaded region, between A and B in this example. We target 90\% of the unit interval in our benchmark with Beta-distributed treatments. }
\end{figure}

\paragraph{Scoring the coverages.}
A reasonable goal would be to achieve a certain amount of coverage~\citep{ref:mccandless} of the true APOs, like having 90\% of the curve be contained in the ignorance intervals. 
Since violation factor $\Gamma$ is not entirely interpretable, nor commensurable across sensitivity models, we measure the size of an ignorance interval via a cost incurred in terms of actionable inference. For each point $t$ of the dose-response curve, we integrated the KL divergence of the actual APO (which defines the $Y_t$ Bernoulli parameter) against the predicted APO uniformly between the bounds. This way, each additional unit of ignorance interval is weighed by its information-theoretic approximation cost. This score is a \emph{divergence cost} of a target coverage. 

\paragraph{Analysis.}
The main results are displayed in Table~\ref{tab:benchmark}. There were ten confounders and the true dose-response curve was a random quadratic form in the treatment and confounders. Other settings are shown in Supplementary~Table~4.
Each trial exhibited completely new projections and outcome function. There were different levels and types of confounding as well as varying model fits. Still, clear patterns are evident in Table~\ref{tab:benchmark}, like the rate at which the $\delta$MSM provided the lowest divergence cost against the baselines. 



\begin{figure}[!ht]\centering 
  \scalebox{0.75}{
    \input{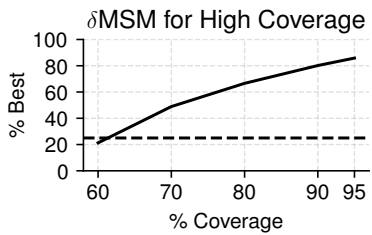}}\vspace{-1em}
  \caption{\label{fig:coverage}Performance for different coverages. Black line: rate of $\delta$MSM achieving lowest divergence cost compared to baselines. Dashed line: expected rate if the chance of any one method outperforming another were identical.}
\end{figure}
\section{A Real-world Exemplar}\label{sec:result-workflow}

The UK Biobank~\citep{ref:bycroft} is a large, densely phenotyped epidemiological study with brain imaging. We preprocessed 40 attributes, eight of which were continuous diet quality scores (DQSs)~\citep{ref:said, ref:zhuang} valued 0--10 and serving as treatments, on 42,032 people. The outcome was thicknesses of 34 cortical brain regions. A poor DQS could translate to noticeable atrophy in the brain of some older individuals, depending on their attributes~\citep{ref:gu,ref:melo}.

\begin{figure}[!hb]
  \scalebox{0.75}{
    \input{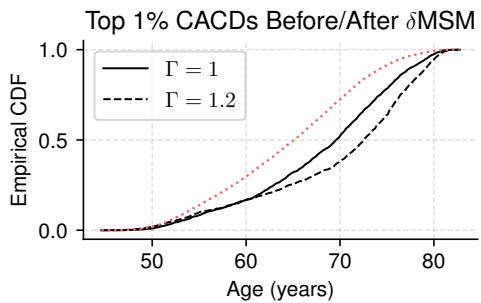}}\vspace{-1em}
  \caption{\label{fig:biobank-age}When we apply the $\delta$MSM $(\Gamma>1)$ for partial identification, the individuals with the top 1\% causal derivatives of cortical thickness with respect to DQSs skew even older. This is expected logically because older people have more years during which they could have revised their diets. Red dotted line corresponds to the entire population.}
\end{figure}

Continuous treatments enable the (Conditional) Average Causal Derivative, {(C)ACD $\coloneqq \partial\E[Y_t|X]\,/\,\partial t$}. The CACD informs investigators on the incremental change in outcome due to a small change in an individual's given treatment. For instance, it may be useful to identify the individuals who would benefit the most from an incremental improvement in diet. We plotted the age distributions of the top 1\% individuals by CACD (diet $\to$ cortical thickness) in Figure~\ref{fig:biobank-age}.

We also compared the $\delta$MSM to an equivalent binary MSM where CACDs are computed in the latter case by thresholding the binary propensity at $t$. Each model's violation factor $\Gamma$ was set for an equivalent amount ($\sim$30\%) of nonzero CACDs. Under the $\delta$MSM, the DQSs with strongest average marginal benefit ranked as vegetables, whole grains, and then meat, for both females and males. They differed under the binary MSM, with meat, then whole grains as the top for females and dairy, then refined grains as the top for males.



\section{Discussion}\label{sec:discussion}


Sensitivity analyses for hidden confounders can help to guard against erroneous conclusions from observational studies. We generalized the practice to causal dose-response curves, thereby increasing its practical applicability. However, there is no replacement for an actual interventional study, and researchers must be careful to maintain a healthy degree of skepticism towards observational results even after properly calibrating the partially identified effects.

Specifically for Average Potential Outcomes (APOs) via the sample-based algorithm, we demonstrated widespread applicability of the $\delta$MSM in \S\ref{sec:result-benchmark} by showing that it provided tighter ignorance intervals than the recent CMSM and other models for $78\%$ of all trials, notwithstanding the wide variation in scenarios tested. Ablating the approximation in Equation~\ref{eq:approx} and dropping the quadratic term, that percentage falls slightly to $74\%$. Even further, keeping just the constant term results in a large drop to $7\%$. This result suggests that the proposed Taylor expansion (Equation~\ref{eq:approx}) is useful, and that terms of higher order would not give additional value.

We showcased sensical behaviors of the $\delta$MSM in a real observational case study (\S\ref{sec:result-workflow}), e.g.\ how older people would be more impacted by (retroactive) changes to their reported diets. Additionally, the top effectual DQSs appeared more consistent with the $\delta$MSM rather than the binary MSM.

\paragraph{Contrasting the CMSM.}
Another recently proposed sensitivity model for continuous-valued treatments is the CMSM~\citep{ref:jesson22}, which was included in our benchmark, \S\ref{sec:result-benchmark}. Unlike the $\delta$MSM, the CMSM does not always guarantee $\underline{d}\leq 1 \leq \overline{d}$ and therefore $p(y|t,x)$ need not be admissible for $\tilde p(y_t|x)$. For partial identification of the CAPO with importance sampling, the propensity density factors out and does not affect outcome sensitivity under the CMSM. For that implementation it happens that $p(y|t,x)$ is indeed admissible. However, we believe that the nominal propensity should play a role in the CAPO's sensitivity to hidden confounders, as both the CMSM and the $\delta$MSM couple the hidden confounding (via the complete propensity) to the nominal propensity. Equations~\ref{eq:expectation}~\&~\ref{eq:expectation-simplified} make it clear that the propensity plays a key role in outcome sensitivity under the $\delta$MSM for both CAPO and APO. We remind the reader of the original MSM that bounds a ratio of complete and nominal propensity odds. The $\delta$MSM takes that structure to the infinitesimal limit and maintains the original desirable property of $p(y|t,x)$ admissibility for $\tilde p(y_t|x)$.

\paragraph{Looking ahead.}
Alternatives to sampling-based Algorithm~\ref{alg:minimax} deserve further investigation for computing ignorance intervals on expectations---but not only. Our analytical solutions bound the density function $p(y_t|x)$ of conditional potential outcomes, which can generate other quantities of interest~\citep{ref:kallus22} or play a role in larger pipelines. Further, an open challenge with the $\delta$MSM would be to find a pragmatic solution to sharp partial identification. Recent works have introduced sharpness to binary-treatment sensitivity analysis~\citep{ref:oprescu23}.



\section{Conclusion}
We recommend the novel $\delta$MSM for causal sensitivity analyses with continuous-valued treatments. The simple and practical Monte Carlo estimator for the APO and CAPO (Algorithm~\ref{alg:minimax}) gives tighter ignorance intervals with the $\delta$MSM than alternatives. We believe that the partial identification of the potential-outcome density shown in Equation~\ref{eq:expectation-simplified}, in conjunction with the parametric formulas of Table~\ref{tab:trusts}, is of general applicability for causal inference in real-world problems. The variety of settings presented in that table allow a domain-informed selection of realistic sensitivity assumptions. For instance, when estimating the effect of a real-valued variable's deviations from some base value, like a region's current temperature compared to its historical average, the \textsf{Gaussian} scheme could be used. \textsf{Gamma} is ideal for one-sided or unidirectional deviations. Finally, \textsf{Balanced Beta} is recommended for measurements in an interval where neither of the endpoints is special.

\begin{acknowledgements}
  This work was funded in part by Defense Advanced Research Projects Agency (DARPA) and Army Research Office (ARO) under Contract No.\ W911NF-21-C-0002.
\end{acknowledgements}

\bibliography{refs}

\begin{thebibliography}{61}
\providecommand{\natexlab}[1]{#1}
\providecommand{\url}[1]{\texttt{#1}}
\expandafter\ifx\csname urlstyle\endcsname\relax
  \providecommand{\doi}[1]{doi: #1}\else
  \providecommand{\doi}{doi: \begingroup \urlstyle{rm}\Url}\fi

\bibitem[Athey et~al.(2019)Athey, Tibshirani, and Wager]{ref:athey19}
S.~Athey, J.~Tibshirani, and S.~Wager.
\newblock Generalized random forests.
\newblock \emph{The Annals of Statistics}, 47\penalty0 (2):\penalty0
  1148--1178, 2019.

\bibitem[Atroszko(2019)]{ref:atroszko2019}
P.~A. Atroszko.
\newblock Is a high workload an unaccounted confounding factor in the relation
  between heavy coffee consumption and cardiovascular disease risk?
\newblock \emph{The American Journal of Clinical Nutrition}, 110\penalty0
  (5):\penalty0 1257--1258, 2019.

\bibitem[Bonvini and Kennedy(2022)]{ref:bonvini}
M.~Bonvini and E.~H. Kennedy.
\newblock Fast convergence rates for dose-response estimation.
\newblock \emph{arXiv preprint arXiv:2207.11825}, 2022.

\bibitem[Bromiley(2003)]{ref:bromiley}
P.~Bromiley.
\newblock Products and convolutions of gaussian probability density functions.
\newblock \emph{Tina-Vision Memo}, 3\penalty0 (4):\penalty0 1, 2003.

\bibitem[Bycroft et~al.(2018)Bycroft, Freeman, Petkova, Band, Elliott, Sharp,
  Motyer, Vukcevic, Delaneau, O’Connell, et~al.]{ref:bycroft}
C.~Bycroft, C.~Freeman, D.~Petkova, G.~Band, L.~T. Elliott, K.~Sharp,
  A.~Motyer, D.~Vukcevic, O.~Delaneau, J.~O’Connell, et~al.
\newblock The uk biobank resource with deep phenotyping and genomic data.
\newblock \emph{Nature}, 562\penalty0 (7726):\penalty0 203--209, 2018.

\bibitem[Calabrese and Baldwin(2001)]{ref:calabrese}
E.~J. Calabrese and L.~A. Baldwin.
\newblock U-shaped dose-responses in biology, toxicology, and public health.
\newblock \emph{Annual Review of Public Health}, 22\penalty0 (1):\penalty0
  15--33, 2001.
\newblock \doi{10.1146/annurev.publhealth.22.1.15}.
\newblock PMID: 11274508.

\bibitem[Chen et~al.(2022)Chen, Minorics, and Janzing]{ref:chen}
Y.-L. Chen, L.~Minorics, and D.~Janzing.
\newblock Correcting confounding via random selection of background variables.
\newblock \emph{arXiv preprint arXiv:2202.02150}, 2022.

\bibitem[Chernozhukov et~al.(2017)Chernozhukov, Chetverikov, Demirer, Duflo,
  Hansen, Newey, Robins, et~al.]{ref:chernozhukov17}
V.~Chernozhukov, D.~Chetverikov, M.~Demirer, E.~Duflo, C.~Hansen, W.~Newey,
  J.~Robins, et~al.
\newblock Double/debiased machine learning for treatment and causal parameters.
\newblock Technical report, 2017.

\bibitem[Chernozhukov et~al.(2021)Chernozhukov, Cinelli, Newey, Sharma, and
  Syrgkanis]{ref:chernozhukov}
V.~Chernozhukov, C.~Cinelli, W.~Newey, A.~Sharma, and V.~Syrgkanis.
\newblock Long story short: Omitted variable bias in causal machine learning.
\newblock \emph{arXiv preprint arXiv:2112.13398}, 2021.

\bibitem[Colangelo and Lee(2021)]{ref:colangelo}
K.~Colangelo and Y.-Y. Lee.
\newblock Double debiased machine learning nonparametric inference with
  continuous treatments.
\newblock \emph{arXiv preprint arXiv:2004.03036}, 2021.

\bibitem[Cornfield et~al.(1959)Cornfield, Haenszel, Hammond, Lilienfeld,
  Shimkin, and Wynder]{ref:cornfield}
J.~Cornfield, W.~Haenszel, E.~C. Hammond, A.~M. Lilienfeld, M.~B. Shimkin, and
  E.~L. Wynder.
\newblock Smoking and lung cancer: recent evidence and a discussion of some
  questions.
\newblock \emph{Journal of the National Cancer institute}, 22\penalty0
  (1):\penalty0 173--203, 1959.

\bibitem[Cristali and Veitch(2022)]{ref:cristali}
I.~Cristali and V.~Veitch.
\newblock Using embeddings for causal estimation of peer influence in social
  networks.
\newblock In A.~H. Oh, A.~Agarwal, D.~Belgrave, and K.~Cho, editors,
  \emph{Advances in Neural Information Processing Systems}, 2022.

\bibitem[Curth et~al.(2021)Curth, Svensson, Weatherall, and van~der
  Schaar]{ref:curth}
A.~Curth, D.~Svensson, J.~Weatherall, and M.~van~der Schaar.
\newblock Really doing great at estimating {CATE}? a critical look at {ML}
  benchmarking practices in treatment effect estimation.
\newblock In \emph{Thirty-fifth Conference on Neural Information Processing
  Systems Datasets and Benchmarks Track (Round 2)}, 2021.

\bibitem[Dorn and Guo(2022)]{ref:dorn22}
J.~Dorn and K.~Guo.
\newblock Sharp sensitivity analysis for inverse propensity weighting via
  quantile balancing.
\newblock \emph{Journal of the American Statistical Association}, pages 1--13,
  2022.

\bibitem[Ghassami et~al.(2021)Ghassami, Sani, Xu, and Shpitser]{ref:ghassami}
A.~Ghassami, N.~Sani, Y.~Xu, and I.~Shpitser.
\newblock Multiply robust causal mediation analysis with continuous treatments.
\newblock \emph{arXiv preprint arXiv:2105.09254}, 2021.

\bibitem[Godos et~al.(2020)Godos, Tieri, Ghelfi, Titta, Marventano, Lafranconi,
  Gambera, Alonzo, Sciacca, Buscemi, et~al.]{ref:godos2020}
J.~Godos, M.~Tieri, F.~Ghelfi, L.~Titta, S.~Marventano, A.~Lafranconi,
  A.~Gambera, E.~Alonzo, S.~Sciacca, S.~Buscemi, et~al.
\newblock Dairy foods and health: an umbrella review of observational studies.
\newblock \emph{International Journal of Food Sciences and Nutrition},
  71\penalty0 (2):\penalty0 138--151, 2020.

\bibitem[Gu et~al.(2015)Gu, Brickman, Stern, Habeck, Razlighi, Luchsinger,
  Manly, Schupf, Mayeux, and Scarmeas]{ref:gu}
Y.~Gu, A.~M. Brickman, Y.~Stern, C.~G. Habeck, Q.~R. Razlighi, J.~A.
  Luchsinger, J.~J. Manly, N.~Schupf, R.~Mayeux, and N.~Scarmeas.
\newblock Mediterranean diet and brain structure in a multiethnic elderly
  cohort.
\newblock \emph{Neurology}, 85\penalty0 (20):\penalty0 1744--1751, 2015.

\bibitem[Guo et~al.(2022)Guo, Yin, Wang, and Jordan]{ref:guo}
W.~Guo, M.~Yin, Y.~Wang, and M.~Jordan.
\newblock Partial identification with noisy covariates: A robust optimization
  approach.
\newblock In \emph{Conference on Causal Learning and Reasoning}, pages
  318--335. PMLR, 2022.

\bibitem[Hill(2011)]{ref:hill}
J.~L. Hill.
\newblock Bayesian nonparametric modeling for causal inference.
\newblock \emph{Journal of Computational and Graphical Statistics}, 20\penalty0
  (1):\penalty0 217--240, 2011.

\bibitem[Hoover et~al.(2020)Hoover, Portillo-Wightman, Yeh, Havaldar, Davani,
  Lin, Kennedy, Atari, Kamel, Mendlen, et~al.]{ref:hoover}
J.~Hoover, G.~Portillo-Wightman, L.~Yeh, S.~Havaldar, A.~M. Davani, Y.~Lin,
  B.~Kennedy, M.~Atari, Z.~Kamel, M.~Mendlen, et~al.
\newblock Moral foundations twitter corpus: A collection of 35k tweets
  annotated for moral sentiment.
\newblock \emph{Social Psychological and Personality Science}, 11\penalty0
  (8):\penalty0 1057--1071, 2020.

\bibitem[Hu et~al.(2021)Hu, Wu, Zhang, and Wu]{ref:hu2021}
Y.~Hu, Y.~Wu, L.~Zhang, and X.~Wu.
\newblock A generative adversarial framework for bounding confounded causal
  effects.
\newblock In \emph{Proceedings of the AAAI Conference on Artificial
  Intelligence}, volume~35, pages 12104--12112, 2021.

\bibitem[Huang et~al.(2021)Huang, Linton, and Zhang]{ref:huang}
W.~Huang, O.~Linton, and Z.~Zhang.
\newblock A unified framework for specification tests of continuous treatment
  effect models.
\newblock \emph{Journal of Business \& Economic Statistics}, 0\penalty0
  (0):\penalty0 1--14, 2021.
\newblock \doi{10.1080/07350015.2021.1981915}.

\bibitem[Jesson et~al.(2020)Jesson, Mindermann, Shalit, and Gal]{ref:jesson20}
A.~Jesson, S.~Mindermann, U.~Shalit, and Y.~Gal.
\newblock Identifying causal-effect inference failure with uncertainty-aware
  models.
\newblock \emph{Advances in Neural Information Processing Systems},
  33:\penalty0 11637--11649, 2020.

\bibitem[Jesson et~al.(2021)Jesson, Mindermann, Gal, and Shalit]{ref:jesson21}
A.~Jesson, S.~Mindermann, Y.~Gal, and U.~Shalit.
\newblock Quantifying ignorance in individual-level causal-effect estimates
  under hidden confounding.
\newblock \emph{ICML}, 2021.

\bibitem[Jesson et~al.(2022)Jesson, Douglas, Manshausen, Solal, Meinshausen,
  Stier, Gal, and Shalit]{ref:jesson22}
A.~Jesson, A.~R. Douglas, P.~Manshausen, M.~Solal, N.~Meinshausen, P.~Stier,
  Y.~Gal, and U.~Shalit.
\newblock Scalable sensitivity and uncertainty analyses for causal-effect
  estimates of continuous-valued interventions.
\newblock In A.~H. Oh, A.~Agarwal, D.~Belgrave, and K.~Cho, editors,
  \emph{Advances in Neural Information Processing Systems}, 2022.
\newblock URL \url{https://openreview.net/forum?id=PzI4ow094E}.

\bibitem[Kallus(2022)]{ref:kallus22}
N.~Kallus.
\newblock Treatment effect risk: Bounds and inference.
\newblock In \emph{2022 ACM Conference on Fairness, Accountability, and
  Transparency}, pages 213--213, 2022.

\bibitem[Kallus and Santacatterina(2019)]{ref:kallus-weighting}
N.~Kallus and M.~Santacatterina.
\newblock Kernel optimal orthogonality weighting: A balancing approach to
  estimating effects of continuous treatments.
\newblock \emph{arXiv preprint arXiv:1910.11972}, 2019.

\bibitem[Kallus et~al.(2019)Kallus, Mao, and Zhou]{ref:kallus}
N.~Kallus, X.~Mao, and A.~Zhou.
\newblock Interval estimation of individual-level causal effects under
  unobserved confounding.
\newblock In \emph{The 22nd international conference on artificial intelligence
  and statistics}, pages 2281--2290. PMLR, 2019.

\bibitem[Kang et~al.(2018)Kang, Subramaniam, Targ, Nguyen, Maliskova, McCarthy,
  Wan, Wong, Byrnes, Lanata, et~al.]{ref:kang}
H.~M. Kang, M.~Subramaniam, S.~Targ, M.~Nguyen, L.~Maliskova, E.~McCarthy,
  E.~Wan, S.~Wong, L.~Byrnes, C.~M. Lanata, et~al.
\newblock Multiplexed droplet single-cell rna-sequencing using natural genetic
  variation.
\newblock \emph{Nature biotechnology}, 36\penalty0 (1):\penalty0 89--94, 2018.

\bibitem[Kilbertus et~al.(2020)Kilbertus, Kusner, and Silva]{ref:kilbertus2020}
N.~Kilbertus, M.~J. Kusner, and R.~Silva.
\newblock A class of algorithms for general instrumental variable models.
\newblock \emph{Advances in Neural Information Processing Systems},
  33:\penalty0 20108--20119, 2020.

\bibitem[Lim et~al.(2021)Lim, Ji, Oberst, Blecker, Horwitz, and
  Sontag]{ref:lim}
J.~Lim, C.~X. Ji, M.~Oberst, S.~Blecker, L.~Horwitz, and D.~Sontag.
\newblock Finding regions of heterogeneity in decision-making via expected
  conditional covariance.
\newblock \emph{Advances in Neural Information Processing Systems},
  34:\penalty0 15328--15343, 2021.

\bibitem[Lo(1987)]{ref:lo}
A.~Y. Lo.
\newblock A large sample study of the bayesian bootstrap.
\newblock \emph{The Annals of Statistics}, 15\penalty0 (1):\penalty0 360--375,
  1987.

\bibitem[Louizos et~al.(2017)Louizos, Shalit, Mooij, Sontag, Zemel, and
  Welling]{ref:louizos}
C.~Louizos, U.~Shalit, J.~M. Mooij, D.~Sontag, R.~Zemel, and M.~Welling.
\newblock Causal effect inference with deep latent-variable models.
\newblock \emph{Advances in neural information processing systems}, 30, 2017.

\bibitem[Manski(2003)]{ref:manski}
C.~F. Manski.
\newblock \emph{Partial identification of probability distributions}, volume~5.
\newblock Springer, 2003.

\bibitem[Mastouri et~al.(2021)Mastouri, Zhu, Gultchin, Korba, Silva, Kusner,
  Gretton, and Muandet]{ref:mastouri}
A.~Mastouri, Y.~Zhu, L.~Gultchin, A.~Korba, R.~Silva, M.~Kusner, A.~Gretton,
  and K.~Muandet.
\newblock Proximal causal learning with kernels: Two-stage estimation and
  moment restriction.
\newblock In \emph{International Conference on Machine Learning}, pages
  7512--7523. PMLR, 2021.

\bibitem[McCandless et~al.(2007)McCandless, Gustafson, and
  Levy]{ref:mccandless}
L.~C. McCandless, P.~Gustafson, and A.~Levy.
\newblock Bayesian sensitivity analysis for unmeasured confounding in
  observational studies.
\newblock \emph{Statist Med}, 26:\penalty0 2331--2347, 2007.

\bibitem[Melo Van~Lent et~al.(2022)Melo Van~Lent, Gokingco, Short, Yuan,
  Jacques, Romero, DeCarli, Beiser, Seshadri, Himali, et~al.]{ref:melo}
D.~Melo Van~Lent, H.~Gokingco, M.~I. Short, C.~Yuan, P.~F. Jacques, J.~R.
  Romero, C.~S. DeCarli, A.~S. Beiser, S.~Seshadri, J.~J. Himali, et~al.
\newblock Higher dietary inflammatory index scores are associated with brain
  mri markers of brain aging: Results from the framingham heart study offspring
  cohort.
\newblock \emph{Alzheimer's \& Dementia}, 2022.

\bibitem[Meresht et~al.(2022)Meresht, Syrgkanis, and Krishnan]{ref:meresht}
V.~B. Meresht, V.~Syrgkanis, and R.~G. Krishnan.
\newblock Partial identification of treatment effects with implicit generative
  models.
\newblock In A.~H. Oh, A.~Agarwal, D.~Belgrave, and K.~Cho, editors,
  \emph{Advances in Neural Information Processing Systems}, 2022.
\newblock URL \url{https://openreview.net/forum?id=8cUGfg-zUnh}.

\bibitem[Mokhberian et~al.(2020)Mokhberian, Abeliuk, Cummings, and
  Lerman]{ref:mokhberian}
N.~Mokhberian, A.~Abeliuk, P.~Cummings, and K.~Lerman.
\newblock Moral framing and ideological bias of news.
\newblock In \emph{Social Informatics: 12th International Conference, SocInfo
  2020, Pisa, Italy, October 6--9, 2020, Proceedings 12}, pages 206--219.
  Springer, 2020.

\bibitem[Oprescu et~al.(2023)Oprescu, Dorn, Ghoummaid, Jesson, Kallus, and
  Shalit]{ref:oprescu23}
M.~Oprescu, J.~Dorn, M.~Ghoummaid, A.~Jesson, N.~Kallus, and U.~Shalit.
\newblock B-learner: Quasi-oracle bounds on heterogeneous causal effects under
  hidden confounding.
\newblock \emph{arXiv preprint arXiv:2304.10577}, 2023.

\bibitem[Padh et~al.(2022)Padh, Zeitler, Watson, Kusner, Silva, and
  Kilbertus]{ref:padh2022}
K.~Padh, J.~Zeitler, D.~Watson, M.~Kusner, R.~Silva, and N.~Kilbertus.
\newblock Stochastic causal programming for bounding treatment effects.
\newblock \emph{arXiv preprint arXiv:2202.10806}, 2022.

\bibitem[Rosenbaum(2002)]{ref:rosenbaum}
P.~R. Rosenbaum.
\newblock \emph{Observational Studies}.
\newblock Springer, 2002.

\bibitem[Rosenbaum and Rubin(1983)]{ref:rosenbaum83}
P.~R. Rosenbaum and D.~B. Rubin.
\newblock Assessing sensitivity to an unobserved binary covariate in an
  observational study with binary outcome.
\newblock \emph{Journal of the Royal Statistical Society: Series B
  (Methodological)}, 45\penalty0 (2):\penalty0 212--218, 1983.

\bibitem[Rubin(1974)]{ref:rubin}
D.~B. Rubin.
\newblock Estimating causal effects of treatments in randomized and
  nonrandomized studies.
\newblock \emph{Journal of Educational Psychology}, 66\penalty0 (5):\penalty0
  688, 1974.

\bibitem[Said et~al.(2018)Said, Verweij, and van~der Harst]{ref:said}
M.~A. Said, N.~Verweij, and P.~van~der Harst.
\newblock Associations of combined genetic and lifestyle risks with incident
  cardiovascular disease and diabetes in the uk biobank study.
\newblock \emph{JAMA cardiology}, 3\penalty0 (8):\penalty0 693--702, 2018.

\bibitem[Sarvet and Stensrud(2022)]{ref:sarvet}
A.~L. Sarvet and M.~J. Stensrud.
\newblock Without commitment to an ontology, there could be no causal
  inference.
\newblock \emph{Epidemiology}, 33\penalty0 (3):\penalty0 372--378, 2022.

\bibitem[Simpson(1951)]{ref:simpson}
E.~H. Simpson.
\newblock The interpretation of interaction in contingency tables.
\newblock \emph{Journal of the Royal Statistical Society: Series B
  (Methodological)}, 13\penalty0 (2):\penalty0 238--241, 1951.

\bibitem[Taleb(2018)]{ref:taleb}
N.~N. Taleb.
\newblock (anti) fragility and convex responses in medicine.
\newblock In \emph{Unifying Themes in Complex Systems IX: Proceedings of the
  Ninth International Conference on Complex Systems 9}, pages 299--325.
  Springer, 2018.

\bibitem[Tan(2006)]{ref:tan}
Z.~Tan.
\newblock A distributional approach for causal inference using propensity
  scores.
\newblock \emph{Journal of the American Statistical Association}, 101\penalty0
  (476):\penalty0 1619--1637, 2006.

\bibitem[Tchetgen et~al.(2020)Tchetgen, Ying, Cui, Shi, and Miao]{ref:tchetgen}
E.~J.~T. Tchetgen, A.~Ying, Y.~Cui, X.~Shi, and W.~Miao.
\newblock An introduction to proximal causal learning.
\newblock \emph{arXiv preprint arXiv:2009.10982}, 2020.

\bibitem[Tokdar and Kass(2010)]{ref:tokdar}
S.~T. Tokdar and R.~E. Kass.
\newblock Importance sampling: A review.
\newblock \emph{WIREs Computational Statistics}, 2\penalty0 (1):\penalty0
  54--60, 2010.

\bibitem[T\"ubbicke(2022)]{ref:tubbicke}
S.~T\"ubbicke.
\newblock Entropy balancing for continuous treatments.
\newblock \emph{J Econ Methods}, 11\penalty0 (1):\penalty0 71--89, 2022.

\bibitem[Vegetabile et~al.(2021)Vegetabile, Griffin, Coffman, Cefalu, Robbins,
  and McCaffrey]{ref:vegetabile}
B.~G. Vegetabile, B.~A. Griffin, D.~L. Coffman, M.~Cefalu, M.~W. Robbins, and
  D.~F. McCaffrey.
\newblock Nonparametric estimation of population average dose-response curves
  using entropy balancing weights for continuous exposures.
\newblock \emph{Health Services and Outcomes Research Methodology}, 21\penalty0
  (1):\penalty0 69--110, 2021.

\bibitem[Veitch and Zaveri(2020)]{ref:veitch}
V.~Veitch and A.~Zaveri.
\newblock Sense and sensitivity analysis: Simple post-hoc analysis of bias due
  to unobserved confounding.
\newblock In H.~Larochelle, M.~Ranzato, R.~Hadsell, M.~F. Balcan, and H.~Lin,
  editors, \emph{Advances in Neural Information Processing Systems}, volume~33,
  pages 10999--11009. Curran Associates, Inc., 2020.

\bibitem[Yadlowsky et~al.(2020)Yadlowsky, Namkoong, Basu, Duchi, and
  Tian]{ref:yadlowski}
S.~Yadlowsky, H.~Namkoong, S.~Basu, J.~Duchi, and L.~Tian.
\newblock Bounds on the conditional and average treatment effect with
  unobserved confounding factors.
\newblock \emph{arXiv preprint arXiv:1808.09521}, 2020.

\bibitem[Yao et~al.(2021)Yao, Chu, Li, Li, Gao, and Zhang]{ref:yao}
L.~Yao, Z.~Chu, S.~Li, Y.~Li, J.~Gao, and A.~Zhang.
\newblock A survey on causal inference.
\newblock \emph{ACM Transactions on Knowledge Discovery from Data (TKDD)},
  15\penalty0 (5):\penalty0 1--46, 2021.

\bibitem[Yin et~al.(2021)Yin, Shi, Wang, and Blei]{ref:yin}
M.~Yin, C.~Shi, Y.~Wang, and D.~M. Blei.
\newblock Conformal sensitivity analysis for individual treatment effects.
\newblock \emph{arXiv preprint arXiv:2112.03493v2}, 2021.

\bibitem[Ystrom et~al.(2022)Ystrom, Degerud, Tesli, H{\o}ye,
  Reichborn-Kjennerud, and N{\ae}ss]{ref:ystrom2022}
E.~Ystrom, E.~Degerud, M.~Tesli, A.~H{\o}ye, T.~Reichborn-Kjennerud, and
  {\O}.~N{\ae}ss.
\newblock Alcohol consumption and lower risk of cardiovascular and all-cause
  mortality: the impact of accounting for familial factors in twins.
\newblock \emph{Psychological Medicine}, pages 1--9, 2022.

\bibitem[Yule(1903)]{ref:yule}
G.~U. Yule.
\newblock {NOTES ON THE THEORY OF ASSOCIATION OF ATTRIBUTES IN STATISTICS}.
\newblock \emph{Biometrika}, 2\penalty0 (2):\penalty0 121--134, 02 1903.
\newblock ISSN 0006-3444.
\newblock \doi{10.1093/biomet/2.2.121}.
\newblock URL \url{https://doi.org/10.1093/biomet/2.2.121}.

\bibitem[Zhao et~al.(2019)Zhao, Small, and Bhattacharya]{ref:zhao}
Q.~Zhao, D.~S. Small, and B.~B. Bhattacharya.
\newblock Sensitivity analysis for inverse probability weighting estimators via
  the percentile bootstrap.
\newblock \emph{Journal of the Royal Statistical Society (Series B)},
  81\penalty0 (4):\penalty0 735--761, 2019.

\bibitem[Zhuang et~al.(2021)Zhuang, Liu, Li, Wan, Wu, Wu, Zhang, and
  Jiao]{ref:zhuang}
P.~Zhuang, X.~Liu, Y.~Li, X.~Wan, Y.~Wu, F.~Wu, Y.~Zhang, and J.~Jiao.
\newblock Effect of diet quality and genetic predisposition on hemoglobin a1c
  and type 2 diabetes risk: gene-diet interaction analysis of 357,419
  individuals.
\newblock \emph{Diabetes Care}, 44\penalty0 (11):\penalty0 2470--2479, 2021.

\end{thebibliography}

\newpage

\appendix
\onecolumn

\section{Completing the Derivations}

Consider Equation~\ref{eq:decompose}.A:
\begin{multline}\label{eq:decompose-further}
  \int_0^1 w_t(\tau) \tilde p(y_t|\tau,x)p(\tau|x)\diff\tau
  = \underbrace{p(y_t|t,x)\int_0^1 w_t(\tau) p(\tau|x)\diff\tau}_{(A.0)} \\
  +\quad \underbrace{g_1(y_t|t,x)\int_0^1 w_t(\tau) (\tau-t) p(\tau|x) \diff\tau}_{(A.1)}
  \quad+\quad \underbrace{g_2(y_t|t,x)\int_0^1 w_t(\tau) \frac{(\tau-t)^2}{2}  p(\tau|x) \diff\tau}_{(A.2)},\\
  \textrm{where}\quad g_k(y_t|t,x)\coloneqq \partial^k_\tau p(y_t|\tau,x)|_{\tau=t}.
\end{multline}
Lightening the notation with a shorthand for the weighted expectations, $\langle \cdot \rangle_\tau \coloneqq \int_{\,0}^1 w_t(\tau) (\cdot) p(\tau|x)\diff\tau,$ it becomes apparent that we must grapple with the pseudo-moments $\langle 1 \rangle_\tau$, $\langle \tau-t \rangle_\tau$, and $\langle (\tau-t)^2 \rangle_\tau$. Note that $t$ should not be mistaken for a ``mean'' value.

Furthermore, we have yet to fully characterize $g_k(y_t|t,x)$. Observe that
\begin{align*}
  p(y_t|\tau,x) = \frac{p(\tau|y_t,x)p(y_t|x)}{p(\tau|x)} \quad&\iff\quad \partial_\tau p(y_t|\tau,x) = p(y_t|x)\cdot\frac{\partial}{\partial\tau} \frac{p(\tau|y_t,x)}{p(\tau|x)}.\\
\intertext{The $p(y_t|x)$ will be moved to the other side of the equation as needed; by Equation~\ref{eq:big-lambda},}
  \frac{\partial}{\partial\tau} \frac{p(\tau|y_t,x)}{p(\tau|x)}
  &= \frac{\partial}{\partial\tau} \Lambda(\tau|y_t,x).\\
\intertext{Expanding,}
  &= \frac{\partial}{\partial\tau}\exp{\int_0^\tau \gamma(\tau|y_t,x) \diff \tau}
  \quad=\quad \gamma(\tau|y_t,x) \exp{\int_0^\tau \gamma(\tau|y_t,x) \diff \tau}\\
  &= (\gamma\Lambda)(\tau|y_t,x).
\end{align*}

Appropriate bounds will be calculated for $g_2(y_t|t,x)$ next, utilizing the finding above as their main ingredient. 
Let
\begin{equation*}
  \tilde g_k(y_t|t,x) \coloneqq p(y_t|x)^{-1}g_k(y_t|t,x)
  = \left.\left(\frac{\partial}{\partial\tau}\right)^{\!k} \frac{p(\tau|y_t,x)}{p(\tau|x)}\right|_{\tau=t.}
\end{equation*}

The second derivative may be calculated in terms of the ignorance quantities $\gamma, \Lambda$:
\begin{align*}
  \tilde g_2(y_t|t,x) =& \partial_\tau \gamma(\tau|y_t,x)\Lambda(\tau|y_t,x)\\
  =& \gamma(\tau|y_t,x)^2 \Lambda(\tau|y_t,x) + \dot\gamma(\tau|y_t,x)\Lambda(\tau|y_t,x)\\
  =& (\gamma^2 + \dot\gamma)\Lambda(\tau|y_t,x).
\end{align*}

And finally we address $\tilde p(y_t|x)$. Carrying over the components of Equation~\ref{eq:decompose-further} into Equation~\ref{eq:decompose},
\begin{equation}\begin{aligned}
  \tilde p(y_t|x) &= \frac{p(y_t|t,x)\langle1\rangle_\tau}{\langle\Lambda(\tau|y_t,x)\rangle_\tau
    - \tilde g_1(y_t|t,x)\langle\tau-t\rangle_\tau - \tilde g_2(y_t|t,x)\langle(\tau-t)^2\rangle_\tau}\\
  &= \frac{p(y_t|t,x)}{\E_\tau[\Lambda(\tau|y_t,x)]
      - (\gamma\Lambda)(t|y_t,x)\E_\tau[\tau-t] - \frac{1}{2}((\dot\gamma+\gamma^2)\Lambda)(t|y_t,x)\E_\tau[(\tau-t)^2]},
\end{aligned}\end{equation}
where these expectations $\E_\tau[\cdot]$ are with respect to the implicit distribution $q(\tau|t,x)\propto w_t(\tau)p(\tau|x).$ The first term in the denominator, $\E_\tau[\Lambda(\tau|y_t,x)]$, may be approximately bounded by the same Algorithm~\ref{alg:minimax}.

\section{How to Calibrate the Weighing Scheme}

We present an argument based on the absolute error of the approximation in Equation~\ref{eq:approx}, specifically for Beta propensities. The following applies to both {\sf Beta} and {\sf Balanced Beta}, $0<t<1$.


Suppose that the the second derivative employed in the Taylor expansion is $Q$-Lipschitz, so that $\abs{\partial^3_\tau p(y_t|\tau,x)} \leq Q.$
Denote the remainder as $\rho(y_t|\tau,x).$ By Taylor's theorem,
\begin{equation*}
\abs{\rho(y_t|\tau,x)} \leq \frac{\abs{\tau-t}^3}{6} Q.
\end{equation*}
The approximated quantity (part A) in Equation~\ref{eq:decompose} is the following integral, which ends up becoming the numerator in Equation~\ref{eq:approx-frac}:
\begin{equation*}
  \int_{0}^1 w_t(\tau)\tilde p(y_t|\tau,x)p(\tau|x)\diff\tau\ =\ \int_{0}^1 w_t(\tau) \big[ p(y_t|\tau,x) + \rho(y_t|\tau,x)\big]p(\tau|x)\diff\tau.
\end{equation*}
The absolute error of this integral is therefore
\begin{equation*}
  \abs{\int_{0}^1 w_t(\tau) \rho(y_t|\tau,x)p(\tau|x)\diff\tau}\ \leq\ \frac{1}{6}Q\underbrace{\int_{0}^1 w_t(\tau)p(\tau|x)\abs{\tau-t}^3\diff\tau}_\text{$\coloneqq J$, which upper-bounds the error.}\quad\text{by the remainder theorem.}
\end{equation*}
Let $A=\alpha-1+rt$ and $B=\beta-1+r(1-t)$, where $(\alpha,\beta)$ parametrize the nominal propensity and $r$ is the precision of the Beta trust-weighing scheme.
The trust-propensity combination is
\begin{equation*}
  w_t(\tau)p(\tau|x) = \frac{\tau^A(1-\tau)^B}{c_t\,\mathbb{B}(\alpha,\beta)},\quad\text{where $c_t=t^{rt}(1-t)^{r(1-t)}$.}
\end{equation*}
Hence, the error bound reduces to
\begin{align*}
  J\ =&\ [c_t\,\mathbb{B}(\alpha,\beta)]^{-1} \int_{0}^1 \tau^A(1-\tau)^B\abs{\tau-t}^3\diff\tau\\
  =&\ [c_t\,\mathbb{B}(\alpha,\beta)]^{-1}\left[\ \underbrace{\frac{\Gamma(A+1)\Gamma(B+1)}{\Gamma(A+B+5)}U_3(A,B,t)}_\text{first term}\ \ +\ \ \underbrace{\frac{\Gamma(A+1)}{\Gamma(A+5)}12t^{A+4}(1-t)^{B+4}\,{_2}F_1(4, A+B+5, A+5;\, t)}_\text{second term}\ \right],
\end{align*}
where $U_3(A,B,t)$ is a cubic polynomial in $A$, $B$, and $t$. Notice that even though the quantity is symmetric about $(A,B,t)\mapsto(B,A,1-t)$, the form does not appear so. We shall focus on the relation of the error bound entirely with $A$ and $\alpha$, then justify the analogous conclusion for $B$ and $\beta$ by the underlying symmetry of the expression.

The Gaussian hypergeometric function in the second term can be expressed as
\begin{align*}
  \sum_{i=0}^\infty \frac{(4)_i(A+B+5)_i}{(A+5)_i} \frac{t^i}{i!}\ =&\ \sum_{i=0}^\infty (4)_i \underbrace{\left(\frac{A+B+5}{A+5}\right)\left(\frac{A+B+6}{A+6}\right)\cdots}_\text{length $i$} \frac{t^i}{i!}\\
  =&\ \sum_{i=0}^\infty \frac{(4)_i}{i!} \left(1+\frac{B}{A+5}\right)\left(1+\frac{B}{A+6}\right)\cdots t^i,\quad\text{where } \frac{(4)_i}{i!} = \frac{(i+2)(i+3)(i+4)}{3!}.
\end{align*}
by using the definition of the Pochhammer symbol $(x)_i=x(x+1)\dots(x+i-1)$. In terms of $A\to\infty$, the whole second term in $J$ is $\mathcal{O}(A^{-4})$ due to the fraction of $\Gamma$ functions. The first term in $J$ is
\begin{equation*}
  \mathcal{O}(A^{-(B+4)}B^{-(A+4)})\cdot U_3(A,B,t)=\mathcal{O}(A^{-B-1}B^{-A-1})
\end{equation*}
by Stirling's approximation of $\Gamma(x)=\mathcal{O}(x^{x-\frac{1}{2}})$.
Clearly, a small $B>0$ might cause the first term in $J$ to explode with large $A$ due to the $\mathcal{O}(B^{-A-1})$ part. This could occur with high $\alpha$, low $\beta$, and low $r$---it is an instance of a high-precision propensity and low-precision weighing scheme destroying the upper error bound. Hence follows an argument for having $r$ match the propensity's precision, to avoid these cases.

As mentioned earlier, the same argument flows for large $B$ and small $A$, while swapping $t\mapsto (1-t).$

\section{Correctness of Algorithm~\ref{alg:minimax}}
The algorithm functions by incrementally reallocating mass (relative, in the weights) to the righthand side, from a cursor beginning on the lefthand side of the ``tape''.
\begin{proof}
  Firstly we characterize the indicator quantity $\Delta_j.$ Differentiate the quantity to be maximized with respect to $w_j;$
  \begin{align*}
    \frac{\partial}{\partial w_j} \frac{\sum_i w_i f_i}{\sum_i w_i} =& \frac{f_j}{\sum_i w_i} - \frac{\sum_i w_i f_i}{\left(\sum_i w_i\right)^2}\\
    =& \frac{f_j\sum_i w_i - \sum_i w_i f_i}{\left(\sum_i w_i\right)^2}\\
    \propto& \underbrace{\sum_i w_i (f_j-f_i)}_{ \coloneqq \Delta_j } \quad \textrm{up to some positive factor.}
  \end{align*}
  Hence, $\Delta_j$ captures the sign of the derivative.

  We shall proceed with induction. Begin with the first iteration, $j=1.$ No weights have been altered since initialization yet. Therefore we have
  \begin{equation*}
    \Delta_1 = \sum_i \overline{w}_i (f_1 - f_i).
  \end{equation*}
  Since $\forall i,\ f_1\leq f_i$ due to the prior sorting, $\Delta_1$ is either negative or zero. If zero, trivially terminate the procedure as all function values are identical.

  Now assume that by the time the algorithm reaches some $j>1$, all $w_k=\underline{w}_k$ for $1\leq k<j$. In other words,
  \begin{equation*}
    \Delta_j = \sum_{i<j}\underline{w}_i \underbrace{(f_j-f_i)}_{(+)} + \sum_{i>j}\overline{w}_i\underbrace{(f_j-f_i)}_{(-)}.
  \end{equation*}
  Per the algorithm, we would flip the weight $w_j\gets \underline{w}_j$ only if $\Delta_j<0.$ In that case,
  \begin{equation*}
    \sum_{i<j}\underline{w}_i (f_j-f_i) < \sum_{i>j}\overline{w}_i(f_i-f_j), \quad \textrm{where both sides are non-negative.}
  \end{equation*}
  Notice that the above is not affected by the current value of $w_j.$ This update can only increase the current estimate because the derivative remains negative and the weight at $j$ is non-increasing. We \emph{must} verify that the derivatives for the previous weights, indexed at $k<j$, remain negative. Otherwise, the procedure would need to backtrack to possibly flip some weights back up.

  More generally, with every decision for weight assignment, we seek to ensure that the condition detailed above is not violated for any weights that have been finalized. That includes the weights before $j$, and those after $j$ at the point of termination. Returning from this digression, at $k<j$ after updating $w_j$,
  \begin{equation*}
    \Delta_k = \sum_{i\leq j} \underline{w}_i (f_k-f_i) + \sum_{i>j}\overline{w}_i(f_k-f_i).
  \end{equation*}
  To glean the sign of this, we refer to a quantity that we know.
  \begin{align*}
    \sum_{i<j}\underline{w}_i (f_j-f_i) <& \sum_{i>j}\overline{w}_i(f_i-f_j)\\
    \iff \sum_{i\leq j}\underline{w}_i (f_k-f_i) <& \sum_{i>j}\overline{w}_i(f_i-f_j)+\sum_{i\leq j} \underline{w}_i(f_k-f_j)\\
    \iff \underbrace{\sum_{i\leq j}\underline{w}_i (f_k-f_i) + \sum_{i>j}\overline{w}_i(f_k-f_i)}_{\Delta_k} <& \underbrace{\sum_{i>j}\overline{w}_i(f_k-f_j)+\sum_{i\leq j} \underline{w}_i(f_k-f_j)}_{\textrm{negative.}}\\
  \end{align*}
  The remaining fact to be demonstrated is that upon termination, when $\Delta_j\geq 0,$ no other pseudo-derivatives $\Delta_{j'},\ j'>j$ are negative. This must be the case simply because $f_{j'}\geq f_j.$
\end{proof}

\twocolumn
\section{On the introductory illustration}

\begin{figure}[!htb]\centering
  \includegraphics[width=\linewidth]{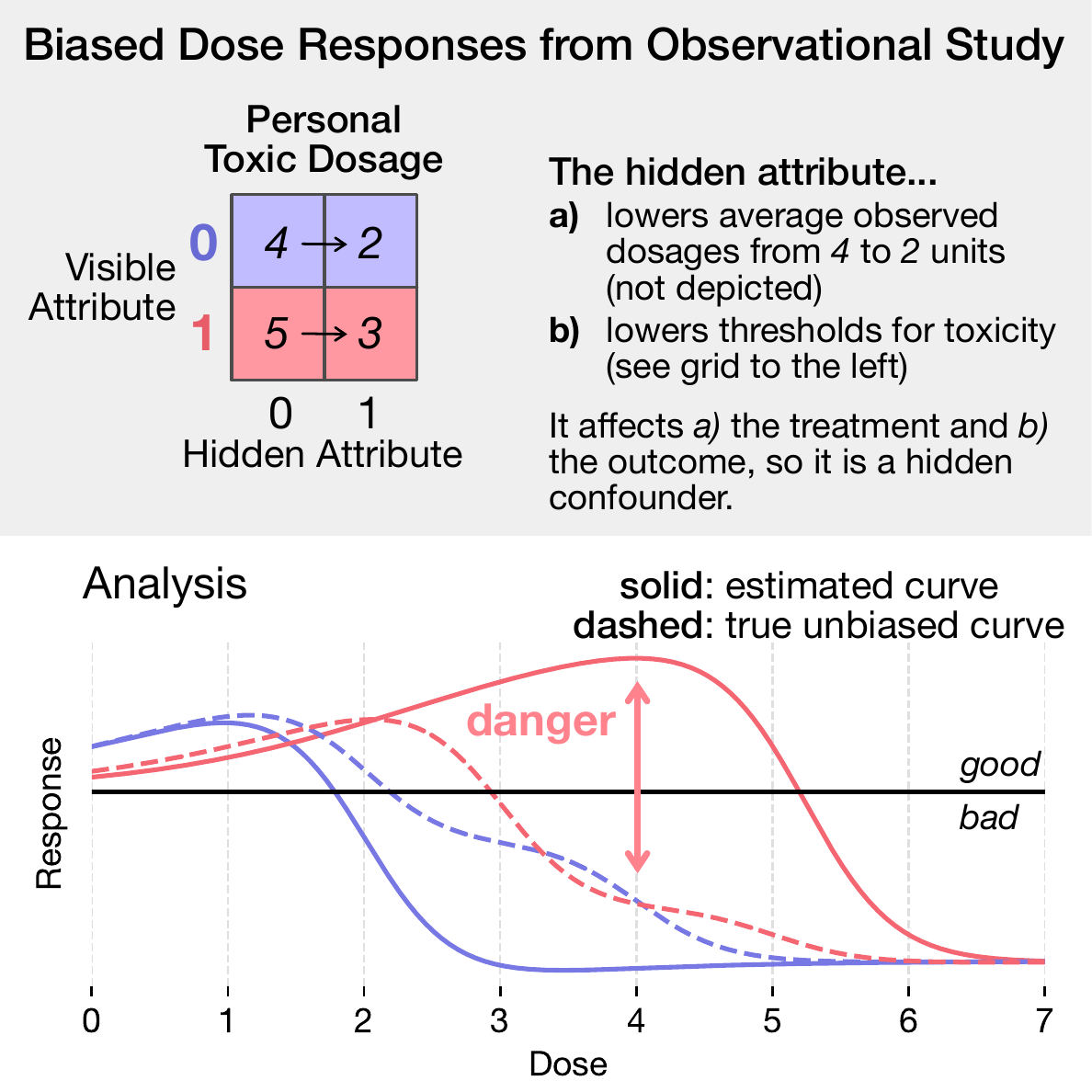}
\caption{\label{fig:curve-flipping-more}
  Elaboration on the example in Figure~\ref{fig:curve-flipping}.
  Treatments were exponentially distributed, and the thresholds displayed in the grid controlled the center of the second sigmoid in $S^2$ due to \citet{ref:taleb}. Two different visible attributes demonstrate how the hidden bias depends on the interplay between propensity and outcome, via the hidden attribute. The blue curve is a little shorter, which allows the vulnerable subgroup's threshold change to be revealed in the data. Estimation minimized the empirical squared error. } 
\end{figure}

\begin{figure}[!hbt]\centering
  \scalebox{0.75}{
    \input{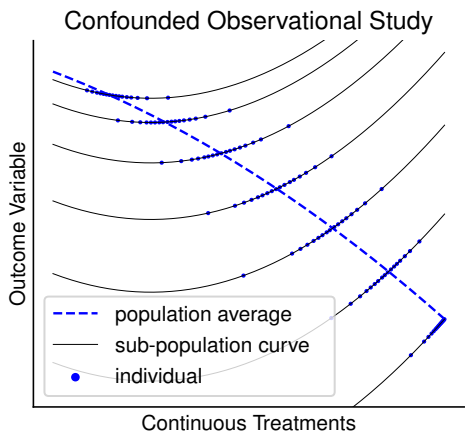}}\vspace{-1em}
\caption{\label{fig:curve-flipping-alt}
  A different example that shows the connection to Simpson's paradox more clearly~\citep{ref:yule,ref:simpson}.
  When a confounder is distorting the assigned treatments in sub-populations, the overall population-level trend may appear flipped in comparison to each sub-population's dose response.}
\end{figure}
\section{Details on The Benchmark}
During each trial, 750 train and 250 test instances of (observed/hidden) confounders, treatment, and outcome were generated. The APO was computed on the test instances. 
Coverage of the dose-response curve was assessed on a treatment grid of $100$ evenly spaced points in $[0,1]$. The different violation factors $\Gamma$ that were tested were also from a $100$-sized grid in $[0,2.5]$.
\begin{align*}
  \intertext{The data-generating process constructed vectors}
  V\coloneqq\langle \text{visible conf\dots, treatment, hidden conf\dots} \rangle\in\mathbb{R}^k
\end{align*}
where $k$ is the number of confounders plus one, for the treatment. Each of these variables is a projection of the original data with i.i.d normal coefficients. We upscale the middle (i.e.\ treatment) entry by $(k-1)$ to keep the treatment effect strong enough. Then, we experiment with two functional forms of confounded dose-response curves:
\begin{itemize}
  \item (linear) mixing vector $\{M_i\}_{i=1}^k \sim \text{i.i.d Normal}(0,1)$. Pre-activation outcome is $u\coloneqq M\cdot v$.
  \item (quadratic) matrix $\{M_{ij}\} \sim \text{i.i.d Normal}(0,1)$. Pre-activation outcome is $u\coloneqq v^\text{T} M v$. Unlike a covariance, $M$ is not positive (semi-)definite. The fact that all entries are i.i.d Gaussian implies that there are cases where the off-diagonal entries are much larger in magnitude than the on-diagonal entries, in such a way that cannot occur in a covariance matrix. This induces more confounding and strengthens our benchmark.
\end{itemize} 

The actual outcome is Bernoulli with probability $u^\star\coloneqq\phi\big((u - m)/s\big)$, wherein $\phi$ is the standard normal CDF, location parameter $m$ is the sample median, and scale $s$ is the sample mean absolute deviation from the median. If $u$ were normal, $s$ would be expected to be a bit smaller than $\sigma$, by a factor of $\sqrt{2/\pi}$. Generally $u^\star$ is no longer uniformly distributed (on margin) because we use $s$, and instead it gravitates towards zero or one. Since the estimated outcome models use logistic sigmoid activations, there is already an intentional measure of model mismatch present in this setup. 

See Table~\ref{tab:benchmark-full} for results under all the settings considered.

The linear outcome and propensity predictors were estimated by maximum likelihood using the ADAM gradient-descent optimizer, with learning rate $10^1$, $4$ batches, and $50$ epochs throughout. For the outcome, we used a sigmoid activation stretched horizontally by $10^2$ for smooth training. For the propensity, similarly, we stretched a sigmoid horizontally and vertically, gating the output in order to yield Beta parameters within $(0,10^2)$.

\paragraph{Data sources.}
The datasets \texttt{brain} and \texttt{blood} both came from the UK Biobank, which is described in the case study of \S\ref{sec:result-workflow}. The two datasets are taken from disjoint subsets of all the available fields, one pertaining to parcelized brain volumes (via MRI) and the other to blood tests. The \texttt{pbmc} dataset came from single-cell RNA sequencing, a modality that is exploding in popularity for bioinformatics. PBMC data are a commonly used benchmark in the field~\citep{ref:kang}. Finally, the \texttt{mftc} dataset consisted of BERT embeddings for morally loaded tweets~\citep{ref:hoover, ref:mokhberian}.

\begin{table}[!ht]\centering
\begin{tabular}{l | r r}
Dataset & Sample Size & Dimension \\
\midrule
\texttt{brain} & 43,069 & 148 \\
\texttt{blood} & 31,811 & 42 \\
\texttt{pbmc} & 14,039 & 16 \\
\texttt{mftc} & 17,930 & 768 \\
\bottomrule
\end{tabular}
\caption{Characteristics of the various datasets employed in our experiments.}
\end{table}

Model mismatch varied with how approximately linear the true dose responses were. As expected, there was a significant negative correlation between model likelihood and divergence cost, so poorer fits had higher costs for coverage. 

\begin{table*}[bt]\centering
  \begin{tabular}{l l l| r r | r r | r r | r r }
    \toprule
    \multicolumn{3}{l}{Benchmarks~{\large\textbackslash}~Scores} & \multicolumn{2}{c}{\tt brain} & \multicolumn{2}{c}{\tt blood} & \multicolumn{2}{c}{\tt pbmc} & \multicolumn{2}{c}{\tt mftc} \\
    & & \multicolumn{1}{l}{} & mean & \multicolumn{1}{r}{median} & mean & \multicolumn{1}{r}{median} & mean & \multicolumn{1}{r}{median} & mean & \multicolumn{1}{r}{median} \\
    \midrule
    linear & 2 confounders & $\delta$MSM & $\bm{94}$ & $\bm{71}$ & $\bm{86}$ & $\bm{63}$ & $\bm{105}$ & $\bm{75}$ & $\bm{69}$ & $\bm{59}$ \\
    & & CMSM & $291$ & $253$ & $261$ & $228$ & $288$ & $259$ & $243$ & $204$ \\
    & & uniform & $116$ & $82$ & $104$ & $71$ & $128$ & $83$ & $78$ & $66$ \\
    & & binary MSM & $116$ & $90$ & $104$ & $73$ & $127$ & $94$ & $91$ & $73$  \\
    \midrule
    & 6 confounders & $\delta$MSM & $\bm{63}$ & $\bm{39}$ & $\bm{63}$ & $\bm{33}$ & $\bm{77}$ & $\bm{44}$ & $\bm{47}$ & $\bm{31}$ \\
    & & CMSM & $177$ & $111$ & $186$ & $117$ & $198$ & $136$ & $167$ & $105$ \\
    & & uniform & $68$ & $41$ & $68$ & $36$ & $83$ & $47$ & $51$ & $33$ \\
    & & binary MSM & $177$ & $176$ & $173$ & $163$ & $188$ & $195$ & $168$ & $160$ \\
    \midrule
    & 10 confounders & $\delta$MSM & $\bm{57}$ & $\bm{31}$ & $\bm{61}$ & $\bm{35}$ & $\bm{72}$ & $\bm{31}$ & $\bm{43}$ & $\bm{27}$ \\
    & & CMSM & $151$ & $81$ & $146$ & $84$ & $158$ & $84$ & $126$ & $74$ \\
    & & uniform & $58$ & $32$ & $63$ & $37$ & $73$ & $33$ & $45$ & $28$ \\
    & & binary MSM & $177$ & $181$ & $182$ & $190$ & $172$ & $170$ & $184$ & $191$ \\
    \midrule
    \ul{quadratic} & 2 confounders & $\delta$MSM & $\bm{170}$ & $\bm{151}$ & $\bm{160}$ & $\bm{139}$ & $\bm{180}$ & $\bm{160}$ & $\bm{159}$ & $\bm{144}$ \\
    & & CMSM & $301$ & $275$ & $283$ & $263$ & $299$ & $274$ & $270$ & $248$ \\
    & & uniform & $198$ & $180$ & $190$ & $166$ & $212$ & $188$ & $190$ & $167$ \\
    & & binary MSM & $205$ & $186$ & $192$ & $169$ & $217$ & $198$ & $190$ & $173$ \\
    \midrule
    & 6 confounders & $\delta$MSM & $\bm{138}$ & $\bm{103}$ & $\bm{145}$ & $\bm{120}$ & $\bm{155}$ & $\bm{134}$ & $\bm{140}$ & $\bm{112}$ \\
    & & CMSM & $216$ & $171$ & $220$ & $193$ & $239$ & $223$ & $222$ & $198$ \\
    & & uniform & $171$ & $118$ & $181$ & $149$ & $189$ & $158$ & $177$ & $132$ \\
    & & binary MSM & $217$ & $231$ & $227$ & $257$ & $230$ & $266$ & $224$ & $249$ \\
    \midrule
    & \ul{10 confounders} & $\delta$MSM & $\bm{138}$ & $\bm{101}$ & $\bm{141}$ & $\bm{100}$ & $\bm{138}$ & $\bm{104}$ & $\bm{144}$ & $\bm{117}$ \\
    & & CMSM & $186$ & $173$ & $188$ & $165$ & $205$ & $178$ & $182$ & $165$ \\
    & & uniform & $158$ & $116$ & $162$ & $108$ & $157$ & $117$ & $167$ & $140$ \\
    & & binary MSM & $211$ & $241$ & $213$ & $240$ & $222$ & $258$ & $214$ & $242$ \\
    \bottomrule
  \end{tabular}
  \caption{\label{tab:benchmark-full}The full array of experiments. Underlined settings are those shown in Table~\ref{tab:benchmark}.}
\end{table*} 


\section{Details on The Biobank Study}
The application number used to access data from the UK Biobank will be mentioned in the de-anonymized manuscript. The measured outcomes were cortical thicknesses and subcortical volumes, the latter normalized by intracranial volume, obtained via structural Magnetic Resonance Imaging (MRI). The results in the main text (\S\ref{sec:result-workflow}) focused on the cortical thicknesses, for brevity. Input variables comprising the covariates and DQS treatments are listed in Table~\ref{tab:biobank-variables}. Inputs were normalized in the unit interval, and outputs were $z$-scored.

\paragraph{Training the models.}
The outcome predictors with 40 inputs and 48 outputs were implemented as multilayer perceptions with three hidden layers of width 32, and single-skip connections. They used Swish activation functions and a unit dropout rate of $0.1$. The ADAM optimizer with learning rate $5\times 10^{-3}$ was was run for $10^4$ epochs. The data were split into four non-overlapping test sets, with separate ensembles of 16 predictors trained for each split. Training sets were bootstrap-resampled for each estimator in the ensemble. The propensity was formulated as a linear model outputting Beta parameters within $(0,64)$, trained in a similar fashion. Finally, CAPOs were partially identified using the set of models from the train-test split for which the data instance belonged to the test set.

\begin{figure}[!ht]\centering
  \includegraphics[width=\linewidth]{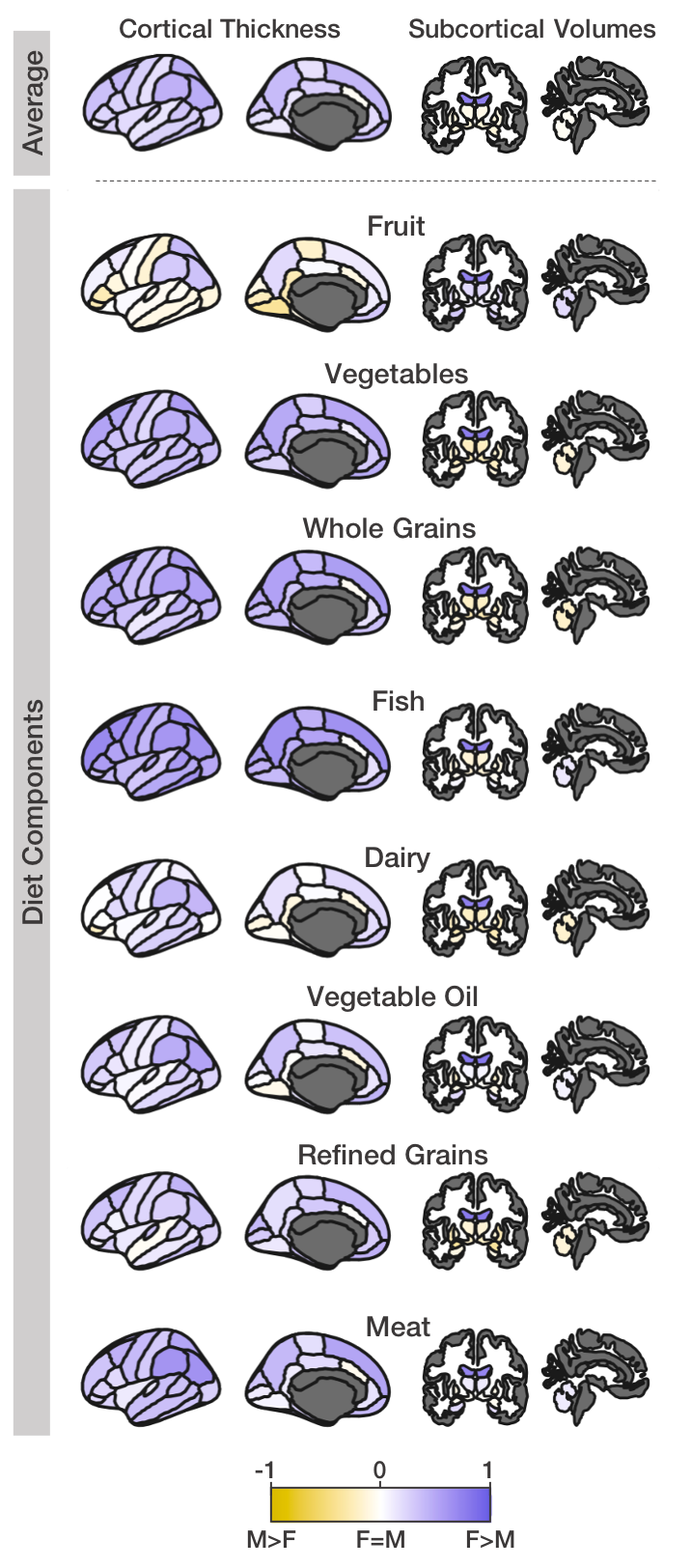}
  \caption{\label{fig:biobank-sexdiff}Normalized effect differences between males and females for the overall average diet score and stratified by individual diet components. The lefthand columns depict individual effects across all cortical thickness parcellations and the righthand side shows subcortical regional volumes. Females show generally larger effects across most diet components.}
\end{figure}

\begin{figure*}[!ht]\centering
  \includegraphics[width=0.85\textwidth]{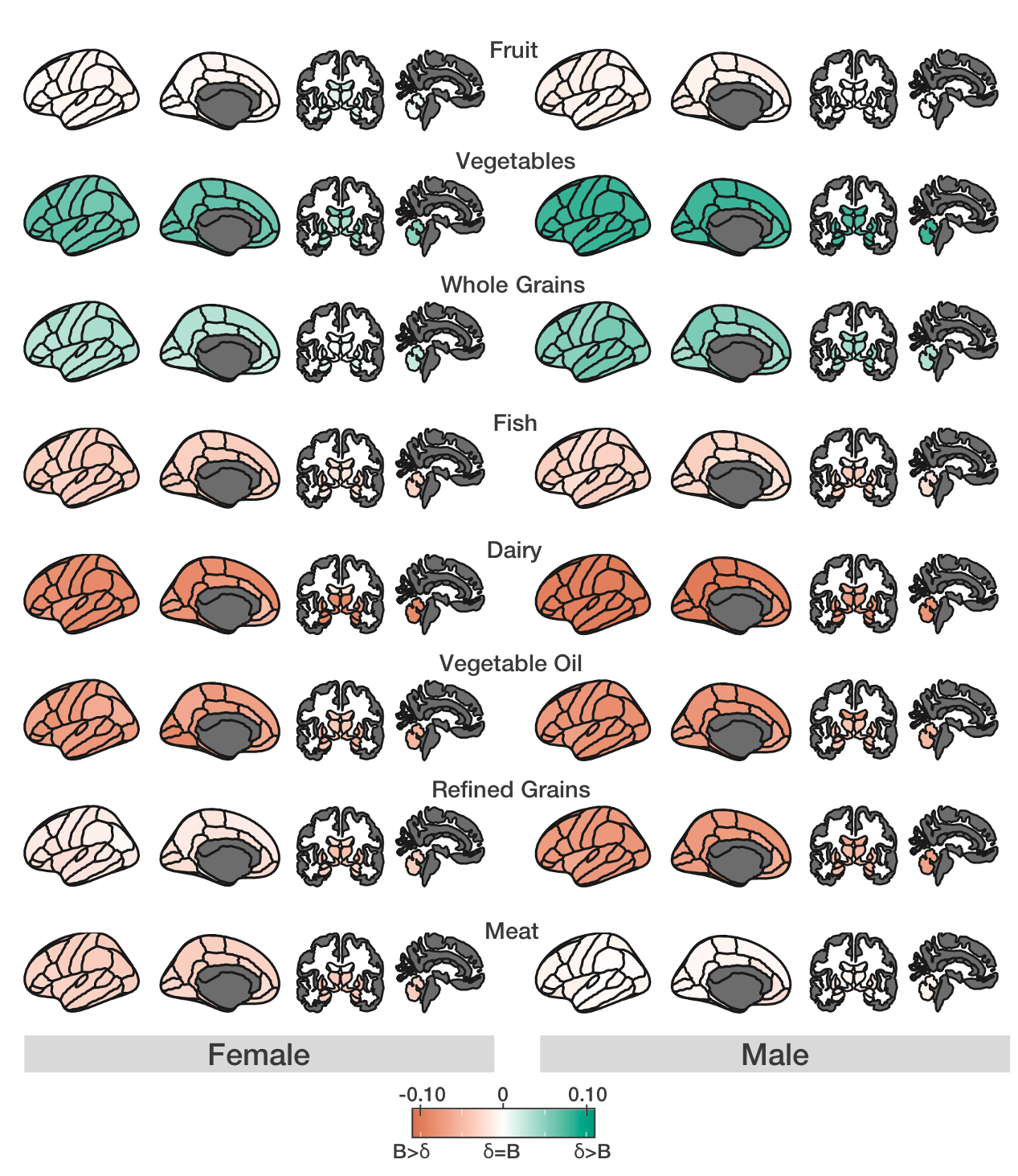}
  \caption{\label{fig:biobank-delta-bin}Normalized effect differences comparing the $\delta$MSM against a shoehorned binary MSM (``$\delta$'' vs. ``B'') stratified by sex. Note differences in relative feature importance, where continuous modeling ranks vegetables and whole grains to be the most important compared to the binary model which emphasizes dairy, vegetable oils, refined grains (primarily for males) and fish.}
\end{figure*}

\paragraph{Additional figures.}
This exploratory study includes plots of relative effects on the various brain regions, shown in Figures~\ref{fig:biobank-sexdiff}~\&~\ref{fig:biobank-delta-bin}. We plan on studying the differential effects of diet on the brain further.

\begin{table*}[bt]
  \small
  \begin{tabular}{|p{1.9cm}|p{4.5cm}|p{6cm}|p{3cm}|}
  \hline
  \textbf{Variable} & \textbf{Features} & \textbf{Classifications} & \textbf{Data Field ID} \\
  \hline
  \multirow{3}{5em}{Demographics} & Age at scan  & - & 21003 \\\cline{2-4}
  & Sex & Male/Female & 31 \\\cline{2-4}
  & Townsend Deprivation Index & - & 189 \\\cline{2-4}
  & ApoE4 copies & 0, 1, 2 & - \\\cline{2-4}
  \hline
  Education& College/University & Yes/No &  6138\\
  \hline
  \multirow{3}{5em}{Physical Activity/ Body Composition} & American Heart Association (AHA) guidelines for weekly physical activity & Ideal ($\geq$150 min/week moderate or $\geq$75 min/wk vigorous or 150 min/week mixed); Intermediate (1--149 min/week moderate or 1--74 min/week vigorous or 1--149 min/week mixed); Poor (not performing any moderate or vigorous activity) & 884, 904,  894, 914 \\\cline{2-4}
  & Waist to Hip Ratio (WHR) & - & 48,49 \\\cline{2-4}
  & Normal WHR & Females: $\le$ 0.85; Males $\le$ 0.90 & 48,49 \\\cline{2-4}
  & Body Mass Index (BMI) & - & 23104 \\\cline{2-4}
  & Body fat percentage & - & 23099 \\\cline{2-4}
  \hline
  \multirow{3}{5em}{Sleep} & Sleep 7-9 Hours a Night &  - & 1160 \\\cline{2-4}
  & Job Involves Night Shift Work & Never/Rarely & 3426 \\\cline{2-4}
  & Daytime Dozing/Sleeping & Never/Rarely & 1220 \\\cline{2-4}
  \hline
  \multirow{3}{5em}{Diet} & DQS 1 - Fruit & - & 1309, 1319 \\\cline{2-4}
  & DQS 2 - Vegetables & - & 1289, 1299 \\\cline{2-4}
  & DQS 3 - Whole Grains & - & 1438, 1448, 1458, 1468 \\\cline{2-4}
  & DQS 4 - Fish & - & 1329, 1339 \\\cline{2-4}
  & DQS 5 - Dairy & - & 1408, 1418 \\\cline{2-4}
  & DQS 6 - Vegetable Oil & - & 1428, 2654, 1438 \\\cline{2-4}
  & DQS 7 - Refined Grains & - & 1438, 1448, 1458, 1468 \\\cline{2-4}
  & DQS 8 - Processed Meats & - & 1349, 3680 \\\cline{2-4}
  & DQS 9 - Unprocessed Meats & - & 1369, 1379, 1389, 3680 \\\cline{2-4}
  & DQS 10 - Sugary Foods/Drinks & - & 6144 \\\cline{2-4}
  & Water intake & Glasses/day & 1528 \\\cline{2-4}
  & Tea intake & Cups/day & 1488 \\\cline{2-4}
  & Coffee intake & Cups/day & 1498 \\\cline{2-4}
  & Fish Oil Supplementation & Yes/No & 20084 \\\cline{2-4}
  & Vitamin/Mineral Supplementation & Multivitamin (with iron/ calcium/ multimineral)/ Vitamins A, B6, B12, C, D, or E/ Folic acid/ Chromium/ Magnesium/ Selenium/ Calcium/ Iron/ Zinc/ Other vitamin & 20084 \\\cline{2-4}
  & Variation in diet & Never/Rarely; Sometimes; Often & 1548 \\\cline{2-4}
  & Salt added to food & Never/Rarely; Sometimes; Usually; Always & 1478 \\\cline{2-4}
  \hline
  Smoking & Smoking status & Never; Previous; Current & 20116\\
  \hline
  Alcohol & Alcohol Frequency & Infrequent (1–3 times a month, special occasions only, or never); Occasional (1–2 a week or 3–4 times a week), Frequent (daily/almost daily and ICD conditions F10, G312, G621, I426, K292, K70, K860, T510) & 1558/ICD\\
  \hline
  \multirow{3}{5em}{Social Support} & Leisure/social activities & Sports club/gym;  pub/social; social/religious; social/adult education; other social group & 6160 \\\cline{2-4}
  & Frequency of Friends/Family Visits & Twice/week or more & 1031 \\\cline{2-4}
  & Able to Confide in Someone & Almost Daily & 2110 \\\cline{2-4}
  \hline
  \end{tabular}
  \caption{\label{tab:biobank-variables}Variables, features, classifications, and respective data fields use in the models.  Diet quality scores (DQS) ranging from 0--10 for 10 components were computed using the same coding scheme as in \citet{ref:said, ref:zhuang}. Leisure/social activity classifications served as their own binary variables. Our results omitted DQS \#8 \& \#10 because they were not even approximately continuous, taking on only a few discrete values.}
\end{table*}

\section{Source-code Availability}
Please visit\ \ \url{https://github.com/marmarelis/TreatmentCurves.jl}.



\end{document}